\documentclass[11pt]{article}

\usepackage{amsmath,amssymb,mathtools,mathrsfs,cite,graphicx,color}
\usepackage{fullpage}
\usepackage[colorlinks=true,urlcolor=blue]{hyperref}
\usepackage{subcaption}
\usepackage{float}
\def \endprf{\hfill {\vrule height6pt width6pt depth0pt}\medskip}

\newcommand{\C}{\mathbb{C}}
\newcommand{\Z}{\mathbb{Z}}

\renewcommand{\Re}[1]{\operatorname{Re}\left\{#1\right\}}
\renewcommand{\Im}[1]{\operatorname{Im}\left\{#1\right\}}

\newcommand{\E}{\operatorname{E}}

\newcommand{\e}{e}
\renewcommand{\j}{j}

\newcommand{\vct}[1]{\boldsymbol{#1}}
\newcommand{\mtx}[1]{\boldsymbol{#1}}
\renewcommand{\H}{\mathrm{H}}
\newcommand{\T}{\mathrm{T}}

\newcommand{\trace}{\operatorname{trace}}

\newcommand{\set}[1]{\mathcal{#1}}
\newcommand{\norm}[1]{\left\lVert#1\right\rVert}

\DeclareMathOperator*{\minimize}{\text{minimize}}
\DeclareMathOperator*{\maximize}{\text{maximize}}

\newcommand{\vh}{\vct{h}}

\newcommand{\vs}{\vct{s}}

\newcommand{\vu}{\vct{u}}
\newcommand{\vv}{\vct{v}}
\newcommand{\vw}{\vct{w}}

\newcommand{\vy}{\vct{y}}
\newcommand{\vz}{\vct{z}}
\newcommand{\valpha}{\vct{\alpha}}

\newcommand{\veta}{\vct{\eta}}
\newcommand{\vtheta}{\vct{\theta}}

\newcommand{\vphi}{\vct{\phi}}

\newcommand{\vpsi}{\vct{\psi}}

\newcommand{\mA}{\mtx{A}}
\newcommand{\mB}{\mtx{B}}
\newcommand{\mC}{\mtx{C}}

\newcommand{\mE}{\mtx{E}}
\newcommand{\mF}{\mtx{F}}

\newcommand{\mL}{\mtx{L}}

\newcommand{\mP}{\mtx{P}}

\newcommand{\mR}{\mtx{R}}

\newcommand{\mT}{\mtx{T}}
\newcommand{\mU}{\mtx{U}}
\newcommand{\mV}{\mtx{V}}
\newcommand{\mW}{\mtx{W}}

\newcommand{\mZ}{\mtx{Z}}
\newcommand{\mGamma}{\mtx{\Gamma}}

\newcommand{\mLambda}{\mtx{\Lambda}}

\newcommand{\mSigma}{\mtx{\Sigma}}

\newcommand{\mPhi}{\mtx{\Phi}}

\newcommand{\mId}{{\bf I}}

\newcommand{\mzero}{{\bf 0}}

\newcommand{\setS}{\set{S}}

\newcommand{\ceil}[1]{\left\lceil #1 \right\rceil}
\newcommand{\inner}[1]{\left<#1\right>}

\begin{document}

\title{Broadband Beamforming via Linear Embedding}
\author{Coleman DeLude, Santhosh Karnik, Mark Davenport, and Justin Romberg}
%\date{DRAFT: \today, \thistime}

\maketitle

\begin{abstract}

In modern applications multi-sensor arrays are subject to an ever-present demand to accommodate signals with higher bandwidths. Standard methods for broadband beamforming, namely digital beamforming and true-time delay, are difficult and expensive to implement at scale. In this work, we explore an alternative method of broadband beamforming that uses a set of linear measurements and a robust low-dimensional signal subspace model. The linear measurements, taken directly from the sensors, serve as a method for dimensionality reduction and serve to limit the array readout.  From these embedded samples, we show how the original samples can be recovered to within a provably small residual error using a Slepian subspace model. 

Previous work in multi-sensor array subspace models have largely analyzed performance from a qualitative or asymptotic perspective.  In contrast, we give quantitative estimates of how well different dimensionality reduction strategies preserve the array gain.  We also show how spatial and temporal correlations can be used to relax the standard Nyquist sampling criterion, how recovery can be achieved through fast algorithms, and how ``hardware friendly'' linear measurements can be designed.
 
\end{abstract}

%--------------------------------------------------------------------------
\section{Introduction}
\label{sec:Introduction}
\begin{figure}[h]
    \centering
    \includegraphics[width=\textwidth]{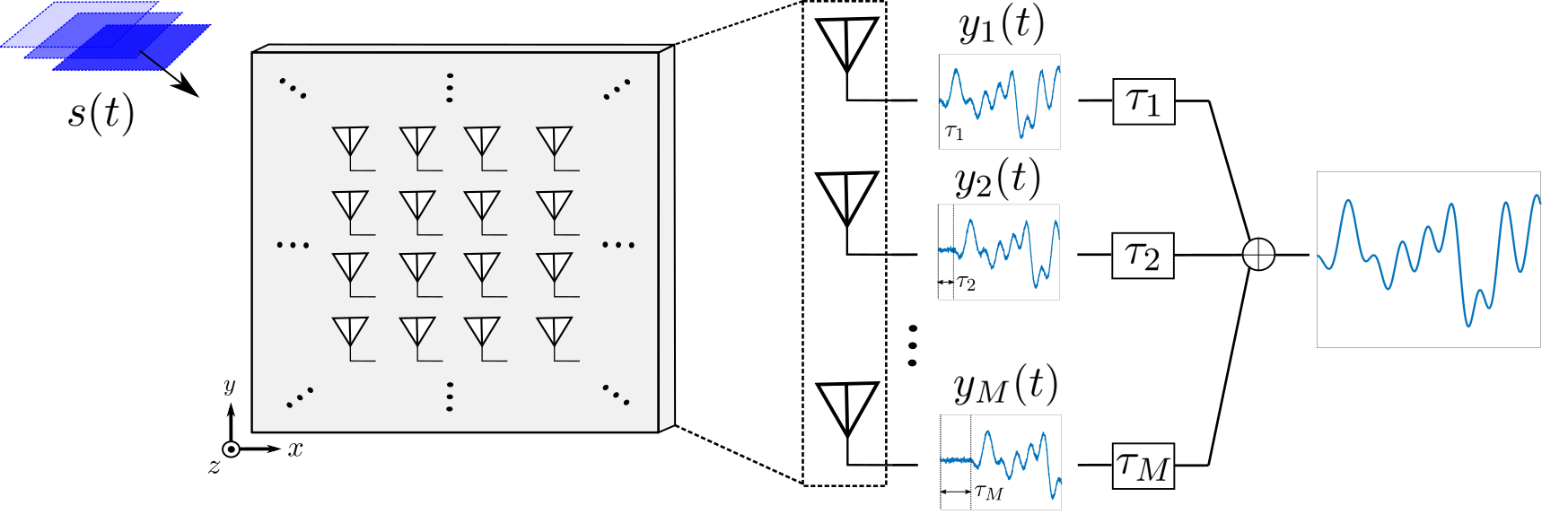}
    \caption{\small \sl Standard array model in which a noisy signal incident to a multi-element array produces a series of time delayed outputs. Beamforming leverages the redundancy in these outputs to reduce noise and provide directional selectivity. For instance, by applying a delay and coherently summing the signals a noise-reduced version of the signal can be produced.}
    \label{fig:Standard_Beamforming}
\end{figure}

This paper revisits the fundamental problem of array processing.  A signal, traveling through space as a plane wave, impinges on a multi-element array (Figure~\ref{fig:Standard_Beamforming}).  We denote what the array center observes as $s(t)$, and the ensemble of outputs from the $M$ array elements as $y_1(t),\ldots,y_M(t)$.  Our model is that each of the $y_m(t)$ will be a different delayed version of $s(t)$,
\begin{align}
    \label{eq:ym}
    y_m(t) = s(t-\tau_m) + \eta_m(t),
\end{align}
where $\tau_m$ is a delay that depends on the relative position of the array element to the array center and the incoming wave (e.g.\ $\tau_m = m(d/c)\cos\theta$ for the linear array in Figure~\ref{fig:Uniform_Samples_ULA}) and $\eta_m(t)$ is noise (or some other perturbation).  Given the output ensemble $\{y_m(t)\}$, we can estimate $s(t)$ by delaying the outputs by the appropriate amount and adding them together,
\begin{align}
	\label{eq:delayandsum}
	\hat{s}(t) &= \sum_{m=1}y_m(t+\tau_m) \\
	\nonumber
	&= Ms(t) + \sum_{m=1}^M\eta_m(t+\tau_m).
\end{align}
This delay-and-sum operation causes the signal parts of each of the outputs $y_m(t)$ to add coherently, while the perturbations add incoherently, thus yielding the array gain.

When the signal $s(t)$ is narrow band, for example say that $s(t) = A\e^{\j\omega_0t}$, then the delay-and-sum in \eqref{eq:delayandsum} amounts to an inner product: delays become multiplies by a unimodular complex number
\begin{equation}
	\label{eq:phaseshift}
	s(t-\tau_m) = \phi_m^* s(t),\quad \phi_m = \e^{\j\omega_0\tau_m},
\end{equation}
and so \eqref{eq:delayandsum} can be replaced by 
\begin{equation}
	\label{eq:weightandsum}
	\hat{s}(t) = \vphi^\H\vy(t),
\end{equation}
where the vector $\vphi\in\C^M$ contains the $\phi_m$ from \eqref{eq:phaseshift} and $\vy(t)\in\C^M$ is a snapshot of the array outputs at time $t$.  As weight-and-sum is typically a much easier operation to execute than delay-and-sum, operations of this type form the conceptual backbone of classical (and modern) beamforming systems.

As is readily seen, the $\vphi$ above depends not only on the direction of arrival (through the $\tau_m$), but also on the frequency $\omega_0$.  If the signal has significant bandwidth\footnote{Equation \eqref{eq:ULA_sub_dim} in Section~\ref{sec:Slepian_Subspace_Model} below makes the notion of ``significant bandwidth'' quantitative for the uniform linear array.}, then \eqref{eq:phaseshift} does not hold, even approximately.  In these cases, \eqref{eq:delayandsum} is implemented in one of two ways: using analog delay lines, or sampling each of the array outputs and implementing the shifts digitally.  Each of these options can be extraordinarily expensive, or even infeasible, as the size of the array get larger and carrier frequencies and bandwidths of the signal increase.

In this paper, we argue that the snapshots $\vy(t)\in\C^M$ of a bandlimited signal $s(t)$ are (essentially) embedded in a low dimensional subspace.  The dimension of this space depends on the signal bandwidth, angle of arrival, and array parameters.  It also depends on the geometry of the array elements, but as we discuss in Section~\ref{sec:Slepian_Subspace_Model} below the dimension can be upper bounded in terms of the \emph{radius} of the array aperture --- for large two-dimensional arrays, the embedding dimension can be significantly smaller than the number of array elements.  The primary consequence of this observation is a third way to estimate $\hat{s}(t)$: the single inner product in \eqref{eq:weightandsum} can be replaced with a matrix-vector multiply, where the matrix has a small number of rows.  This gives us a natural way to extend beamforming to broadband signals, or, equivalently, a simple way to reduce array read-out without sacrificing array gain.

Our framework for broadband beamforming is based on the now classical work of Landau, Pollack, and Slepian on prolate spheroidal wave functions \cite{SlepianI,SlepianII,SlepianIII,SlepianVI,SlepianV}.  As we discuss in detail in Section~\ref{sec:Slepian_Subspace_Model} below, a snapshot $\vy(t)$ can be thought of as a set of discrete samples of a bandlimited signal inside of a limited time interval.  A qualitative upper bound on the number of degrees of freedom of in these samples, which can be derived in this case from the aperture of the array and the bandwidth on the signal, is well-known \cite{karnik2020improved}, implying that $\vy(t)$ is embedded in a low dimensional subspace.  Recent results, based on the careful study of the associated prolate matrix give precise non-asymptotic bounds on this dimension, and have led to fast algorithms for projecting onto ``Slepian spaces'' \cite{karnik2017fast} and estimating bandlimited signals from non-uniform samples \cite{Karnik:2019:nonuniform}.  The main contribution of this paper is to explore the implication of these results in the context of array processing.  In particular,
\begin{itemize}
	\item In Section~\ref{sec:Slepian_Subspace_Model} we give an explicit formulation of the subspace structure for array snapshots ($\vy(t)$ above) of arbitrarily bandlimited signals.  We describe how the energy concentrates in a \emph{Slepian subspace} whose dimension depends on bandwidth, array aperture, and angle of arrival.  Of particular interest is the case of two-dimensional arrays, where we consider both separable and non-separable subspaces for the model, and show that for large arrays and high-bandwidths, the non-separable model is significantly tighter.
	\item In Section~\ref{sec:Measurement_and_recon} we demonstrate how our subspace model can be used to limit array readout by taking a small number of linear measurements, effectively generalizing \eqref{eq:weightandsum}.  We discuss the effectiveness of several strategies (including Slepian projections, forming beams for multiple center frequencies, and random projections) using a standard least-squares formulation and analysis.
	\item In Section~\ref{sec:Temporal_Decimation} we show that our model also allows for a reduction in the temporal sampling frequency to below the traditional Nyquist rate.  Although subsampling may decrease the effective array gain, it gives us another way in which read-out can be naturally reduced.
	\item In Section~\ref{sec:Fast_Slepian_Computations} we discuss fast algorithms for estimating $\hat{s}(t)$ both from this limited readout and from direct measurements.  
	\item In Section~\ref{sec:Signal_Isolation} we give a mathematical characterization of the ``sidelobes'' of our broadband beamformer by bounding the angle between Slepian subspace corresponding to different angles of arrival and different frequency bands.
	\item In Section~\ref{sec:Simplified_Measurements} we present a technique for designing optimal readout projections that are constrained to use weights of $\pm 1$ or $\pm\j$.  The hardware required to implement this type of readout is far easier to implement that a general linear projection while being almost as efficient. 
\end{itemize}

% As a concrete example, discussed in full in Section~\ref{sec:XXX} below, is that a signal with bandwidth $\Omega$ centered at frequency $\omega_0$ impinging on an array of $M$ elements can be captured with $5$ inner products of the form \eqref{eq:weightandsum}.

%--------------------------------------------------------------------------
% SECTION: Related work

\section{Related work}
\label{sec:Related_Work}

The notion of using a signal subspace model to perform beamforming was first popularized by the long celebrated MUSIC algorithm \cite{Schmidt:1986,Spielman:1986}. Though an explicitly narrowband process the foundational idea of forming a signal and noise subspace estimate served as inspiration to find analogous representations in the broadband regime. In \cite{Buckley:1988:BASSALE,Buckley:1986} a broadband source subspace is derived and subsequently used to form MUSIC-like pseudo-spectrums for broadband direction of arrival (DOA) estimation. This subspace is the same as our proposed Slepian space representation, and the authors note several similar properties to those discussed in Section~\ref{sec:Slepian_Subspace_Model}. Further properties of the broadband subspace representation were noted in \cite{Buckley:1987}, with a particular emphasis on the representation being invariant to the array configuration. 
In contrast to these works, this paper is focused on how the Slepian model can be used for acquisition. We give precise estimates on the degree of dimensionality reduction, analysis for using the model for different types of acquisition embeddings (and discuss how these embeddings might be designed), and examine trade-offs for temporal sampling.
%
%However, the justification for the validity of this model is highly reliant on empirical evidence and asymptotic bounds. In contrast our work leverages non-asymptotic bounds to construct a more rigorous and precise mathematical justification. Furthermore our application is centered on signal acquisition as opposed to DOA.

%This property, which is used throughout this paper, was again largely justified by empirical evidence and conjecture on the discrete and continuous time spectral concentration problem. This paper formally justifies this property by leveraging finite sampling bounds that converge to an analogous result on the continuum.

Reduced Dimension BeamSpace (RDBS) is the term generally used to describe the process of finding low dimensional representations of signals received by multi-sensor arrays \cite{Krim:1996}. The motivations for this process are the same as our own; a limited array readout is far easier to handle with respect to several areas of processing. The majority of previous work in this field is centered on a narrowband assumption, with works such as \cite{anderson93op} proving that nearly all the information about multiple narrowband signals at the array can be captured through a linear embedding. Extension to the broadband case requires a more intricate formulation as simple metrics such as beam pattern become somewhat ambiguous. In \cite{Buckley:1987:RDBS} several metrics are proposed for characterizing broadband RDBS performance for use in DOA.  Again, our focus in this paper is how dimensionality reduction can be used for acquisition.
%
%These metrics were largely derived to characterize how broadband DOA was effected by dimensionality reduction. Consequently they differ from our proposed metrics since the motivations behind each are different.

From a structural perspective, the possible use of both analog and digital linear operations in our method resembles hybrid multiple-input multiple-output (MIMO) precoding schemes\cite{Ahmed:2017,Ali:2017}. Such systems leverage a reduced number of RF chains to perform the same task as fully digital precoding, and manifests as a multi-measurement system. We will now examine several works that attempt to solve similar problems to our own, but in the context of MIMO.

% In a similar manner to RDBS processing the notion of leveraging low-dimensional representations (equivalently stated as ``spatial sparsity" in the literature) exists in MIMO precoding as well. The work \cite{Ayach:2014} makes use of a spatial sparsity constraint to design optimal precoding schemes. This work inherently relies on narrowband signals and heavily constrains the choice in subspace to reflect such an assumption. Further work by \cite{Venugopal:2017} leverages spatially sparsity in terms of channel estimation. However, it relies on strict DOA constraints to ensure sparsity and fails to extend to a general non-OFDM signals. As a general statement, it appears that MIMO precoding have had difficulty characterizing spatial sparsity on the continuum.

Since a key feature of our design is a concise method of estimating the subspace dimension and consequently the required number of measurements (each requiring a processing chain) we will now examine works that have investigated similar estimates of this number in MIMO. In \cite{Bogale:2016}, an estimate of the degrees of freedom is formed by analyzing the rank of the composite precoding matrix. However, this formulation is not easily interpretable in terms of parameters such as bandwidth and DOA due to the non-trivial composition of individual precoding matrices. An analogous result was derived in \cite{Sohrabi:2016}, but suffers from the same shortcomings as the previous. In a similar manner to our framework, \cite{Zhu_Li:2017} formulates a ``unified" subspace as well as an easily interpretable bound on the number of necessary processing chains. However, the signals are assumed to be narrowband and the authors explicitly decouple spatial dimensions which leads to a drastic over estimation of the signals underlying degrees of freedom.

There are also a number of recent works that treat broadband beamforming in a fundamentally different way than we do in this paper. For an array tuned to receive a signal from a specified DOA a significant deviation from the assumed center frequency causes an apparent shift in the DOA. This phenomena, termed ``beam-squinting," will cause the signal to fall outside the beam and become attenuated \cite{Jang:2019}. Works such as \cite{Chen:2018,Peng:2018} attempt to remedy this by effectively widening the beam to account for these apparent changes in DOA.  Our approach is to encode the array outputs using multiple linear combinations which might be interpreted as forming several specially designed beams (with associated back-end processing for reconstruction).

The notion of reconstructing samples off the array using a time-domain oriented approach was explored to some degree in \cite{Morsali:2020}. It is shown that leveraging a hybrid precoding architecture one can reconstruct windows of the signal with a reduced number of processing chains. However, this work explicitly leverages an OFDM signal model and consequently fails to generalize to less idealized cases.

A common theme in the previously discussed sources is that the goal of many authors is to approximate a true time-delay. The motivation for this is that actually implementing a time-delay system is extremely difficult for a variety of reasons; namely resolution, power consumption, and chip size \cite{Jung:2020}. Nonetheless a significant amount of research has been put into true time-delay systems in recent years. In \cite{Jang:2019}, a baseband true time delay unit is proposed that is meant to be used in conjunction with phase shifters. Though the inclusion of phase shifters reconciles the issues of RF and baseband delays it adds another level of hardware complexity. The actual time delay operator in this paper is similar to that of \cite{Spoof:2020}. Both sources utilize a re-sampling technique in which the signal is delayed by varying the clock phase of a modified sampling unit. Though not nearly as cumbersome as a tapped delay line, this method still requires a significant amount of additional front-end hardware.

A consequence of our mathematical model is that we can define a precise condition where the narrowband assumption fails. Though this is a minor note in our work and implicitly derived in \cite{Buckley:1986,Buckley:1987,Buckley:1988:BASSALE}, the limits to which a system can handle increased bandwidth is generally not precisely characterized.  A method for determining when the rank of the covariance matrix exceeds one is described in \cite{Zatman:1998}. The approach centers around an analysis of when two sinusoids become sufficiently spaced to produce a rank two signal covariance matrix.  Our approach hinges on the observation that the received bandlimited signal is implicitly time-limited by the array meaning that its covariance matrix will always be full-rank, albeit ill-conditioned to the point where the effective degrees of freedom remain small \cite{SlepianV,karnik2020improved,karnik2017fast}.

\section{Slepian subspace model}
\label{sec:Slepian_Subspace_Model}
To describe our model, we return to Figure~\ref{fig:Standard_Beamforming}.  We will assume the signal impinging upon the array $s(t)$ is a stationary, ergodic, centered, complex, Gaussian random process with power spectral density (PSD) that is $2\Omega$ bandlimited and resides about a center frequency $f_c$,
\begin{align}
\label{eq:GRP_flat_PSD}
S(f) = \begin{cases} 
1, & f \in [f_c-\Omega,f_c+\Omega], \\ 
0, & \text{otherwise.}
\end{cases}
\end{align}
For the purposes of discussion in this section, we will assume that the noise $\veta$ in \eqref{eq:ym} is zero, so that $\vy(t) = \{s(t-\tau_m)\}_m$ is exactly a collection of samples of $s(t)$ at different delays.

\begin{figure}[h]
	\centering
	\begin{subfigure}[b]{0.32\textwidth}
		\includegraphics[width = \textwidth]{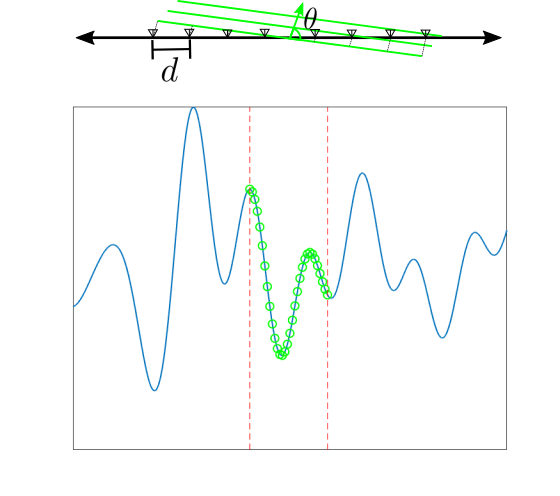}
		\caption{}
	\end{subfigure}
	\begin{subfigure}[b]{0.32\textwidth}
		\includegraphics[width = \textwidth]{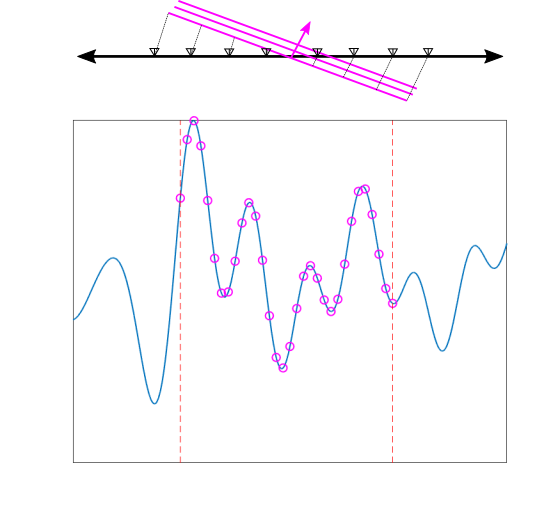}
		\caption{}
	\end{subfigure}
	\begin{subfigure}[b]{0.32\textwidth}
		\includegraphics[width = \textwidth]{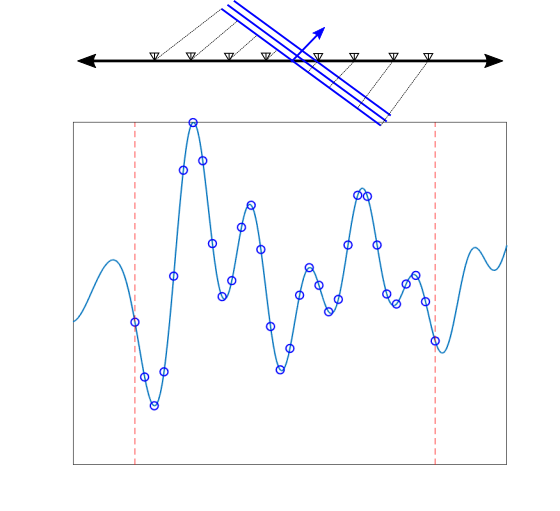}
		\caption{}
	\end{subfigure}
	\caption{\small\sl For a uniform linear array each snapshot of the array produces a set of $M$ uniformly spaced samples of the impinging signal. The close spacing of these samples relative to the Nyquist rate means they are highly correlated, as is apparent from the relative smoothness between samples. The sampling interval and degree of correlation is dependent on angle of arrival with (a) being near broadside (b) at $45^o$ and (c) being near endfire.}
	\label{fig:Uniform_Samples_ULA}
\end{figure}

For the uniform linear array shown in Figure~\ref{fig:Uniform_Samples_ULA} with $s(t)$ incoming at angle $\theta$, a snapshot $\vy(t)$ is a collection of $M$ uniformly spaced samples of $s(t)$ with spacing $\tau = (d/c)\cos\theta$.  If the spacing of the sensors is tuned to the carrier frequency $f_c$, meaning that the sensors are a half wavelength $d=\frac{c}{2f_c}$ apart, and $f_c\gg \Omega$, then $\vy$ corresponds to a ``burst'' of $M$ samples that are much more closely spaced than the Nyquist spacing of $1/2\Omega$.

The close spacing between the samples makes them heavily correlated; the number of effective degrees of freedom in $\vy$ is far below the number of elements $M$.  In fact, the effective degrees of freedom in uniform samples of a bandlimited signal taken over a limited time interval is now well-understood.  The classical work by Landau, Pollack, and Slepian \cite{SlepianI,SlepianII,SlepianIII,SlepianVI,SlepianV} gives a framework for describing these degrees of freedom using the eigenvalue decomposition of the $M\times M$ matrix
\[
\mB[m,n] = \begin{cases} 2W, & m=n, \\ \frac{\sin(2\pi W(m-n))}{\pi(m-n)}, & m\not = n\end{cases}
\]
where $W = \tau\Omega$ (the sample spacing times the bandlimit), which in our uniform linear array example means $W = (\Omega/2f_c)\cos\theta$.  When $s(t)$ is a Gaussian random process as described above, then (with proper normalization) for every $t$, $\vy(t)$ is a Gaussian random vector with covariance matrix 
\[
    \mR = \mE_{f_c}\mB\mE_{f_c}^H,
\]
where $\mE_{f_c}$ is diagonal with $\mE_{f_c}[m,m] = e^{-j2\pi f_c\tau_m}$.

\begin{figure}[h]
        \centering
        \begin{subfigure}[b]{0.32\textwidth}
            \centering
            \includegraphics[width=\textwidth]{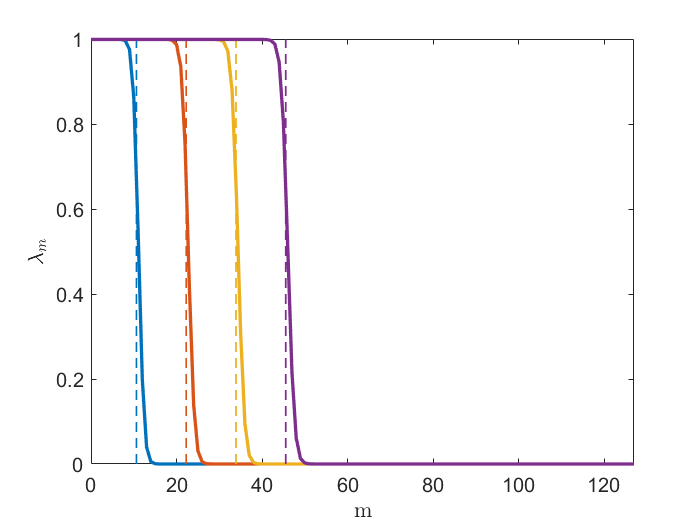}
            \caption%
            {{\small}}    
        \end{subfigure}
        \hfill
        \begin{subfigure}[b]{0.32\textwidth}  
            \centering 
            \includegraphics[width=\textwidth]{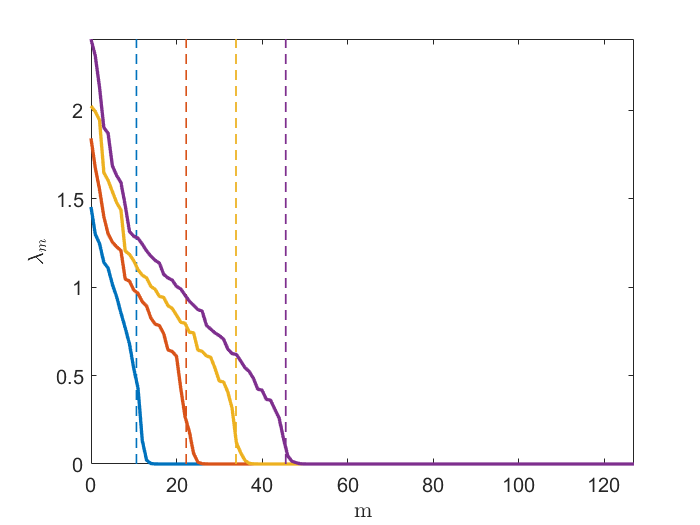}
            \caption%
            {{\small}}    
        \end{subfigure}
        \hfill
        \begin{subfigure}[b]{0.32\textwidth}  
            \centering 
            \includegraphics[width=\textwidth]{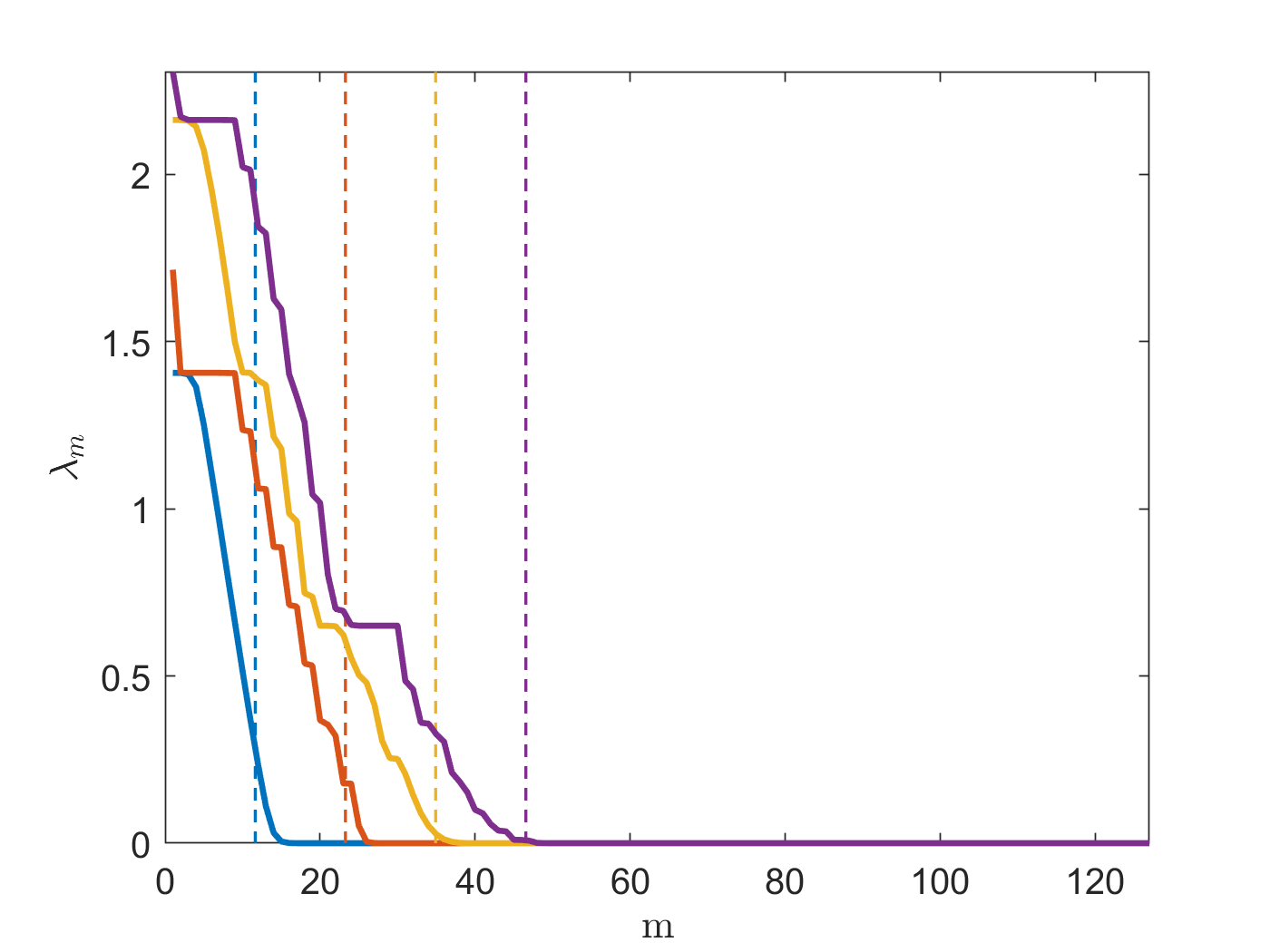}
            \caption%
            {{\small}}    
        \end{subfigure}
                \caption[Projection]
        {\small\sl Plot of the eigenvalue spectrum of a $128 \times 128$ Prolate matrix under a variety of time-bandwidth products for (a) uniform sampling, (b) non-uniform sampling, and (c) samples taken from a $8\times16$ uniform planar array. The dashed lines indicate the time-bandwidth product associated with the eigenvalues of the same color. For the uniform case the the first $2WM$ eigenvalues cluster close to 1 while the eigenvalues exceeding the $2WM$ point cluster close to 0. Eigenvalues that fall outside of these two clusters in the so-called ``transition region" are provably few in number \cite{karnik2017fast}. Though this clean clustering is lost in the non-uniform cases the time-bandwidth product still acts as a threshold for the sharp spectral roll-off.
        % \note{We should consider combining this with Figure 3, as there is some redundancy here.  Also, we need to talk about how the number $128$ relates to the intrinsic quantities $\Omega,T,M$.}
        } 
        \label{fig:prolate_eigenvalue_non_uniform}
\end{figure}

The matrix $\mB$ is approximately, but not exactly, a projection onto a subspace of dimension $2WM$.  An example eigenvalue spectrum of $\mB$ for $f_c=120$ GHz, $M=128$, $\theta = \pi/6$, and $\Omega\in \{12.6,25.2,37.8,50.4\}$ GHz (and so $2WM$ varies) is shown in Figure~\ref{fig:prolate_eigenvalue_non_uniform}(a), where we see that the largest $2WM$ or so eigenvalues of $\mB$ are $\approx 1$ after which the spectrum decays to zero at an extremely fast rate.  What this means is that an array snapshot $\vy(t)$ of a ``typical'' bandlimited signal will have the vast majority of its energy focused in the subspace spanned by the eigenvectors corresponding to the $\lceil 2WM\rceil$ coordinates. This notion is made precise in \eqref{eq:taillambdak} below.
%More precisely, by typical we mean bandlimited signals that that remain spectrally concentrated in-band when viewed over finite sets of samples. Though one can construct bandlimited signals that do not exhibit this behavior they are generally not seen in practice.

Below, we will refer to the eigenvector of $\mB$ corresponding to the $k$th largest eigenvalue as the $k$th {\em Slepian vector}\footnote{Slepian vectors may also be defined as a time-limited version of the discrete prolate spheroidal sequences (DPSS) \cite{SlepianV,Davenport:2011}}, and the span of the first $K$ eigenvectors as the $K$th {\em Slepian subspace}.
We also note that the eigenvalues of $\mR$ are the same as those for $\mB$, while the eigenvectors of $\mR$ can be obtained by applying $\mE_{f_c}$ to the eigenvectors of $\mB$.  As such, we will refer to the eigenvectors of $\mR$ as \emph{modulated Slepian vectors}.

% \begin{figure}[h]
% 	\centering
% 	\includegraphics[width = .5\textwidth]{figs/Prolate_Matrix_Eigenspectrum.png}
% 	\caption{\small\sl {\color{red} [We should index the horizontal axis above by $k$.]} The eigenvalue spectrum of the prolate matrix demonstrates that the first $2WM$ eigenvalues cluster close to 1 while the eigenvalues exceeding the $2WM$ point cluster close to 0. Eigenvalues that fall outside of these two clusters in the so-called ``transition region" are provably few in number \cite{karnik2017fast}.}
% 	\label{fig:Prolate_Matrix_Eigenspectrum}
% \end{figure}

Recent results give a very precise characterization of the eigenvalue spectrum of $\mB$.  In \cite{karnik2020improved}, it was shown that for $L=\lceil 2WM\rceil$, the $k$th largest eigenvalue $\lambda_k$ of $\mB$, with $k\geq L$, satisfies
\begin{equation}
	\label{eq:lambdak}
	\lambda_k ~\leq~ c_1\exp\left(-\frac{k-L}{c_2}\right),\quad k\geq L,
\end{equation}
where $c_1$ and $c_2$ are known (and reasonably small) constants\footnote{See \cite[Cor.\ 1]{karnik2020improved} for a precise statement of \eqref{eq:lambdak} and \eqref{eq:taillambdak}.} that depend on $\log(L+1)$.  That is, the size of the eigenvalues decays exponentially after hitting the edge of the plateau at $k=L$.  This result leads immediately to a quantitative bound on the amount of energy the snapshot $\vy(t)$ will have outside of the $K$th Slepian subspace; if $\vy$ has covariance $\mR$, and we let $\mP_K$ denote the projector onto the $K$th modulated Slepian subspace, then
\begin{equation}
	\label{eq:taillambdak}
	\E\left[\|\vy(t) - \mP_K\vy(t)\|_2^2\right] = \sum_{k=K+1}^M\lambda_k \leq c_3\exp\left(-\frac{K-L}{c_4}\right), \quad K\geq L,
\end{equation}
where again $c_3$ and $c_4$ are constants that depend only on $\log(L+1)$.  Taking $K$ just a little larger than $L$ will ensure that the vast majority of the energy in $\vy(t)$ is focused in the $K$th Slepian subspace. A complementary result to \eqref{eq:taillambdak} from \cite[Lemma 5.3]{Davenport:2011} shows that the energy in a particular realization of $\vy(t)$ will be concentrated in the $K$th Slepian subspace with high probability.
%
% Should we include a figure here that shows the energy concentration for a particular configuration but for multiple randomly generated signals?
%
A reasonable heuristic, then, is that the snapshot $\vy(t)$ across a uniform linear array of a bandlimited source arriving from angle $\theta$ lies (to a close approximation) in a subspace $\setS_\theta^{\text{ULA}}$ of dimension
\begin{align}
	\label{eq:ULA_sub_dim}
	%\vy(t)\in\setS_\theta^{\text{ULA}},\quad 
	\operatorname{dim}(\setS_\theta^{\text{ULA}}) = \max\left(L,1\right), \quad
	L = \ceil{\frac{\Omega}{f_c}M|\cos\theta|}.
\end{align}
Our model for a snapshot $\vy(t)$, in other words, is that it can be closely approximated by a linear combination of $K$ pre-determined basis vectors\footnote{Note that the $\{\vpsi_k\}$ depend on $\theta$ even though we are not making this dependence explicit in our notation.}
\begin{equation}
	\label{eq:yexpansion}
	\vy(t) \approx \sum_{k=1}^K\alpha_k(t)\vpsi_k.
\end{equation}
The (time-varying) snapshot $\vy(t)$ is now characterized by the (also time-varying) lower-dimensional basis coeffcients $\{\alpha_k(t)\}_{k=1}^K$.  There are of course many ways to choose a basis that spans a particular subspace.  For the uniform array here, two natural choices (that have the additional property of orthogonality) is to take the $\{\vpsi_k\}$ to be the first $K$ modulated Slepian vectors (the leading eigenvectors of $\mR$) or a concatenation of discrete Fourier vectors with a small number of auxiliary vectors as studied in  \cite{karnik2017fast,zhu18ro}.

The central concept above extends easily to arbitrary array patterns and arrays arranged in two or three dimensions.  A key feature of the bounds above, and something that was only uncovered in the recent analyses \cite{karnik2020improved,Boulsane:2019,Bonami:2021}, is that they only depend on the product $WM$.  This means that the results can be translated to the continuum (let the number of samples $M$ grow as the space between them $\tau$ shrinks at the same rate, keeping $WM$ constant).  Over an interval of time $T$, \emph{any} set of $M$ samples, no matter their configuration within the interval, can be embedded in a subspace of approximately $2T\Omega$.  It could be that the effective dimension turns out to be much lower, as when the samples are clustered together or when $M<2T\Omega$, but the upper bound is always $2T\Omega$. To illustrate this point, consider the same scenario discussed above to produce Figure~\ref{fig:prolate_eigenvalue_non_uniform}(a) but with the sensors placed uniformly at random over the interval $[0,M\cdot d]$. The temporal lags $\{\tau_m\}_{m=1}^M$ now vary between elements such that when normalized by $\frac{\cos\theta}{2f_c}$ they are no longer integer values. Letting $\tilde\tau_m$ denote the normalized lags we form the generalized prolate matrix
\[
	\tilde\mB[m,n] = \begin{cases} 2W, & \tilde{\tau}_m=\tilde{\tau}_n, \\ \frac{\sin(2\pi W(\tilde{\tau}_m-\tilde{\tau}_n))}{\pi(\tilde{\tau}_m-\tilde{\tau}_n)}, & \tilde{\tau}_m\not = \tilde{\tau}_n\end{cases}.
\]
As was done in the uniform case Figure~\ref{fig:prolate_eigenvalue_non_uniform}(b) displays the eigenspectrum of this matrix for a variety of $2 T \Omega$. It is clear that in both cases the time-bandwidth product acts as a cutoff point for the eigenvalue decay. 
%Hence our estimate of the underlying subspace dimension is invarient under a choice in sensor placement over the aperture. 

\begin{figure}[!htbp]
    \centering
    \includegraphics[width=.5\textwidth]{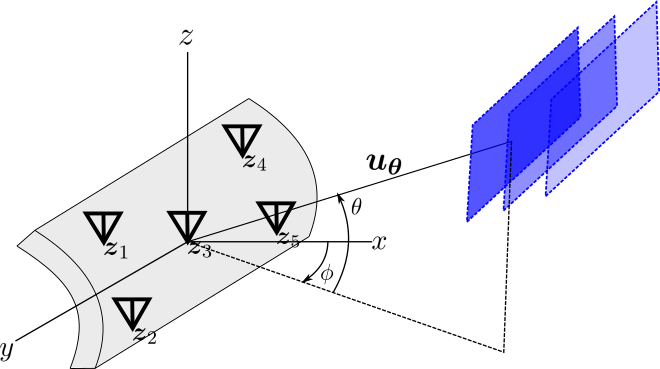}
    \caption{\small \sl Set of sensors $\{\vz_m\}_{m=1}^5$ placed on a cylindrical surface in 3-space with a plane-wave incident at a fixed azimuth and elevation described by the normal vector $\vu_{\vtheta}$. The effective aperture is essentially the depth of the array when viewed from the angle of arrival. }
    \label{fig:Effective_aperture_diagram}
\end{figure}

To see how this translates to our general array processing scenario consider $M$ sensors placed at arbitrary points $\{\vz_m\}_{m=1}^M$ in 3-space, and consider an impinging plane wave whose direction of arrival in terms of azimuth $\phi$ and elevation $\theta$ is denoted as $\vtheta = [\phi, \ \theta]^T$. The normal vector associated with this plane wave is $\vu_{\vtheta} = [\cos \phi \cos\theta,\sin \phi \cos\theta,\sin\theta]^T$, and  a visualization of this scenario for 5 sensors placed on a cylindrical surface is shown in Figure \ref{fig:Effective_aperture_diagram}.
Element $m$ sees a delayed (again relative to the array center) version $s(t-\tau_m)$ of the signal where $\tau_m = \vz_m^\T\vu_{\vtheta}/c$.  The effective aperture of the array is
\begin{align}
	A_{\vtheta} = \left |\max_{m} \vz_m^\T\vu_{\vtheta} - \min_{m} \vz_m^\T\vu_{\vtheta} \right |,
	\label{eq:effective_aperture}
\end{align}
which can be be interpreted as the physical depth of the array as viewed from the angle of arrival, and the width of the temporal window of $s(t)$ impinging on the array at any moment is $T=A_{\vtheta}/c$.  Thus the number of degrees of freedom in a snapshot $\vy(t)$ consisting of samples at the $\{\tau_m\}$ is at most $2T\Omega = 2A_{\vtheta}\Omega/c$.
% this comment is already covered with the discussion above
%(This again is a well-known rule-of-thumb based on classical asymptotics \cite{Slepian:1976} for which quantitative bound of the form \eqref{eq:lambdak},\eqref{eq:taillambdak} have been recently put forth in \cite{karnik2020improved}.)  
No matter how many elements we have and how they are spaced, a snapshot in time will always lie very close to a fixed subspace of dimension at most $2A_{\vtheta}\Omega/c$. (Which subspace this is, however, of course relies critically on the actual element locations.)  In support of this claim Figure~\ref{fig:prolate_eigenvalue_non_uniform}(c) displays the eigenspectrum associated with samples taken along a planar multi-sensor array, which is simply a specific choice in non-uniform sampling pattern. As in Figure~\ref{fig:prolate_eigenvalue_non_uniform}(b) the spectrum exhibits a sharp eigenvalue decay past the $2T\Omega$ threshold. A generalization of  \eqref{eq:ULA_sub_dim} for a bandlimited source impinging on an array with arbitrarily spaced elements from a fixed angle is that the snapshot of samples $\vy(t)$ lies (to a close approximation) in a subspace  $\mathcal{S}_{\vtheta}$ of dimension
\begin{align}
	\label{eq:Arbitruary_sub_dim}
	\text{dim}(\mathcal{S}_{\vtheta}) = \max{\left( \ceil{\frac{2\Omega A_{\vtheta}}{c}},1 \right)}.
\end{align}
%{\color{red} [What is above isn't quite accurate.  It's more like $\vy(t)$ is approximately in the subspace, and the subspace has dimension equal to the quantity on the right.]}  
This again means that our model for $\vy(t)$ is that it is closely approximated by a superposition of  $K = \ceil{\frac{2\Omega A_{\vtheta}}{c}}$ as in \eqref{eq:yexpansion}.  In this case, we can take the $\vpsi_k$ to be samples of the prolate spheroidal wave functions (PSWFs) at the corresponding $\{\tau_m\}$; these vectors would not be orthogonal but in general would form a  well-conditioned basis. 
% We might want a reference for the above. \cite{Karnik:2019:nonuniform}.

Several array geometries admit a closed form expression for \eqref{eq:Arbitruary_sub_dim}. In particular, for an $M\times N$ planar array with element spacing $c/2f_c$ tuned to the carrier frequency, if we assume the angle of arrival is only limited in azimuth (as is generally the case in array design) then the effective aperture given by (\ref{eq:effective_aperture}) reduces to $A_{\vtheta}= \frac{c}{2f_c}\sqrt{M^2 + N^2}|\cos\theta|$. Therefore a snapshot  $\vy$  approximately lies in a subspace $\mathcal{S}_{\theta}^{\text{UPA}}$ with dimension
\begin{align}
	\label{eq:UPA_sub_dim}
	\text{dim}(\mathcal{S}_{\theta}^{\text{UPA}}) = \max{\left( \ceil{\frac{\Omega}{f_c}\sqrt{M^2 + N^2}|\cos\theta|},1 \right)}.
\end{align}
What is interesting about (\ref{eq:UPA_sub_dim}) is that the number of degrees of freedom does not scale with the number of elements $MN$, instead it grows as the far more modest rate $\sqrt{M^2 + N^2}\leq M+N$. This means that for large arrays in two or three space, the underlying subspace dimension can be \emph{drastically} smaller than the number of elements in the array.

\begin{figure}[h]
	\centering
	\includegraphics[width = .75 \textwidth]{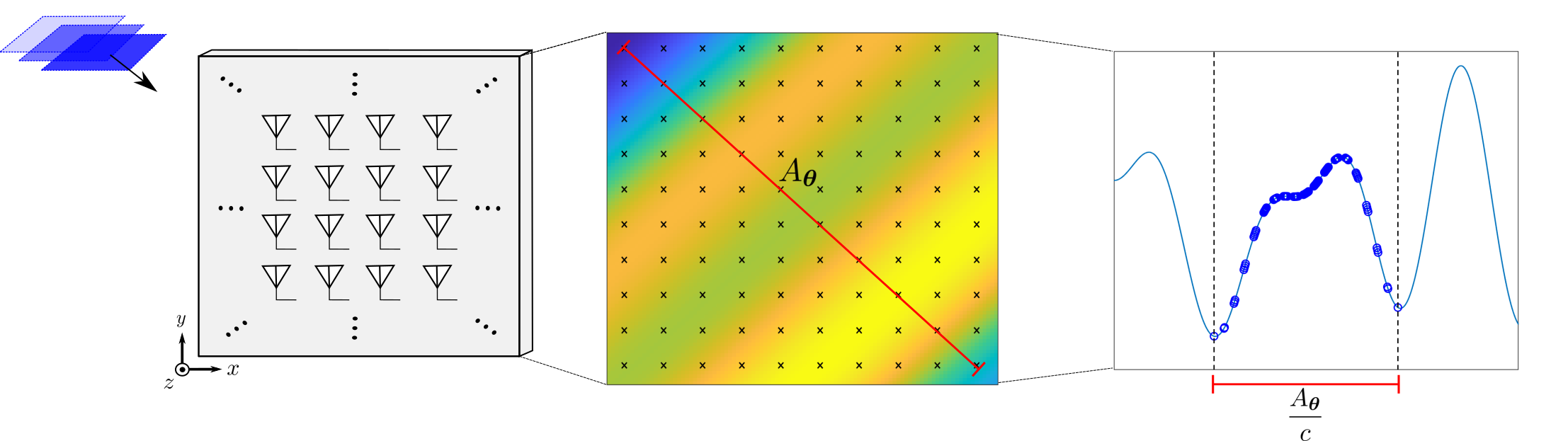}
	\caption{\small\sl For a plane wave incident to a $10\times 10$ array the signal temporally varies across the effective aperture of the array, indicated here by a red line. The signal is then subsequently sampled by the sensors's at the points given by the black x's. Thus a snapshot of the array produces a set of samples from $s(t)$ observed over a temporal window with a width determined by the effective aperture. Different angles of arrival amount to different temporal windows with varying sampling patterns within.}
	\label{fig:Sampling_Pattern_Aperture_vs_Angle}
\end{figure}

An example for a $10\times 10$ uniform planar array (UPA) array is shown in Figure~\ref{fig:Sampling_Pattern_Aperture_vs_Angle} where the impinging signal has a bandwidth of 8.5 GHz and resides at a center frequency of 28 GHz. We see that each snapshot provides a sampling of the signal over a temporal window determined by the effective aperture which in turn is also dependent on the angle of incidence.  Furthermore, the sampling pattern within this temporal window also depends on the angle of arrival. In this example scenario the approximate dimension of the underlying subspace is 4 while the ambient dimension of the UPA array is 100. Hence there are drastically fewer degrees of freedom than the number of array elements as predicted by our previous discussion. 
% {\color{red} [We need to finish the numerical example started above to come up with an actual number here that is less than $100$ ... and we might even show how the dimension grows as the side lengths are increased]}

We note that the representation for the subspace in \eqref{eq:UPA_sub_dim} is inherently \emph{non-separable}.  For the 2D uniform array, we might be tempted to use a separable 2D Slepian representation.  To do this, we could represent each of the $M$ rows in the array using $\approx N\Omega/f_c\cdot |\cos\phi\cos\theta|$ Slepian functions as in \eqref{eq:yexpansion}, then represent these results across the $N$ columns using $\approx M\Omega/f_c|\cos\phi\sin\theta|$ Slepian functions for each.  The result is an embedding in a subspace of dimension 
\[
		\operatorname{dim}(\mathcal{S}_{\vtheta}^{\mathrm{separable}}) \approx 
		\max{\left(\ceil{N\frac{\Omega}{f_c}|\cos\phi\cos\theta|},1 \right)}\cdot
		\max{\left(\ceil{M\frac{\Omega}{f_c}|\cos\phi\sin\theta|},1 \right)}.
\]	
This gives us dimensionality reduction similar to the 1D case in \eqref{eq:ULA_sub_dim}, but is far less than the non-separable representation in \eqref{eq:UPA_sub_dim}.  One way to interpret this savings is that the signal that lies across the 2D array at any moment in time is, as a function of space, is not only spatially bandlimited in 2D wave number space, it is also constant along one direction (as it is a plane wave).  This gives it far more structure than simply being bandlimited, and the non-separable representation above takes advantage of this structure.
\section{Measurement and reconstruction}
\label{sec:Measurement_and_recon}

Our discussion in the previous section establishes a model for a sample snapshot taken across an array on which a bandlimited signal is impinging.  This model was derived from the fact that ``typical'' bandlimited signals will induce a vector of samples whose energy is tightly concentrated in a low dimensional subspace.  If we model the incoming signal as a Gaussian random process with flat power spectral density over a certain band, any snapshot $\vy(t)$ will be a Gaussian random vector with covariance $\mR$ that has a number of significant eigenvalues that depends only on the size of the aperture, the bandwidth of the incoming signal, and (more weakly) on the direction of arrival (recall Figure~\ref{fig:prolate_eigenvalue_non_uniform}).  

In this section, we discuss how a particular snapshot of samples $\vy_0=\{s_m(t_0-\tau_m)\}_{m=1}^M$, for some $t_0 \in \mathbb{R}$, can be recovered from the linear measurements\footnote{We are using a slightly different noise model here than in the Introduction, with the perturbation being added to measurements instead of the signals themselves.  We could account for $\veta$ as in \eqref{eq:ym} by changing the covariance in the analysis below.}
\begin{align}
	\label{eq:measurement_model}
	\vw = \mPhi \vy_0 + \veta,
\end{align}
where $\veta\sim \mathcal{N_C}(\mtx{0},\sigma^2\mId)$ is a benign noise vector and $\mPhi$ is a $K\times M$ matrix.  The action of $\mPhi$ in \eqref{eq:measurement_model} can be viewed as a type of generalized beamforming --- each entry of $\vw$ is a different weighted summation of the entries of $\vy_0$ (as in \eqref{eq:weightandsum}), and so we can interpret each row of $\mPhi$ as defining a different ``beam''.  The main question we are interested in answering is how well this collection of linear measurements ``captures'' $\vy_0$.  We answer this question by analyzing the accuracy a least-squares reconstruction of $\vy_0$ from $\vw$ and how this accuracy depends on the number of rows in $\mPhi$.  We will see that the subspace structure in general allows a number of rows that is significantly smaller than the number of array elements, pointing towards a mechanism for dimensionality reduction at the sensor that can (again significantly) reduce array read-out.

With $\vw$ in hand and the knowledge that $\vy_0\sim \mathcal{N_C}(\mtx{0},\mR)$, we find the minimum mean-square estimate of $\vy_0$ by solving
\begin{align}
    \label{eq:MMSE_recon}
    \minimize_{\vy \in \mathbb{C}^m} \mathbb{E}_{\vy_0,\veta}\left \{\norm{\vy-\vy_0}_2^2 \big| \vw\right \},
\end{align}
where $\mathbb{E}_{\vy_0,\veta}$ denotes the expectation taken over $\vy$ and $\veta$.  The well-known closed form solution to \eqref{eq:MMSE_recon} is\footnote{The expressions \eqref{eq:MSE_sol} and \eqref{eq:MSE} can be derived using the fact that $\begin{bmatrix} \vy_0 \\ \vw \end{bmatrix}$ is a Gaussian random vector with covariance $\begin{bmatrix} \mR & \mR\mPhi^H \\ \mPhi\mR & \mPhi\mR\mPhi^H + \sigma^2\mId \end{bmatrix}$ and then computing the conditional mean and conditional covariance of $\vy_0$ given $\vw$.}
\begin{align}
    \label{eq:MSE_sol}
    \hat\vy = \mR\mPhi^H(\mPhi\mR\mPhi^H+\sigma^2\mtx{I})^{-1}\vw,
\end{align}
which yields a corresponding mean-square error of 
\begin{align}
    \label{eq:MSE}
    \mathbb{E}_{\vy_0,\veta}\left \{\norm{\hat\vy-\vy_0}_2^2 \big| \vw\right \} = \trace{\left (\mR-\mR\mPhi^H(\mPhi\mR\mPhi^H+\sigma^2\mtx{I})^{-1}\mPhi\mR \right )}.
\end{align}
This gives us an expression for evaluating the effectiveness of any given measurement matrix $\mPhi$ for a particular $\mR$.

It is also a classic result that if we restrict the spectral norm (largest singular value) of $\mPhi$ to be $\|\mPhi\|\leq 1$, then the optimal choice of $\mPhi$, the one that makes \eqref{eq:MSE} as small as possible, is to take the rows to be the $K$ leading eigenvectors of $\mR$.  That is, if the covariance has eigenvalue decomposition $\mR = \mV\mLambda\mV^H$, where the (real-valued) entries along the diagonal are sorted from largest to smallest, then we take $\mPhi = \mV_K^H$, where $\mV_K$ consists of the first $K$ columns of $\mV$; a quick proof of this fact is provided in Appendix~\ref{apx:optlinmeas} for completeness.  As we saw in the last section, in the case of a ULA, these eigenvectors are modulated Slepian basis vectors.  For arrays with general geometry, the eigenvectors are an othonormal basis whose span matches that of the first $K$ PSWFs sampled in the appropriate places.

With $\mPhi=\mV_K^H$, \eqref{eq:MSE} reduces to 
\begin{align*}
    \mathbb{E}_{\vy_0,\veta}\left \{\norm{\hat\vy-\vy_0}_2^2 \big| \vw\right \} = \sum_{m=1}^K \lambda_m\left(1-\frac{\lambda_m}{\lambda_m+\sigma^2}\right) + \sum_{m=K+1}^M \lambda_m
\end{align*}
where the multiplicative factors $1-\lambda_m/(\lambda_m+\sigma^2)$ in the first sum are $\approx 0$ for $\lambda_m \gg \sigma^2$ and $\approx 1$ for $\lambda_m \ll \sigma^2$. If the eigenvalue $\lambda_m$ is very small, it makes almost no difference whether it is included in the first summation or the second.  
Given the discussion of the eigenvalue behavior of $\mR$ in Section~\ref{sec:Slepian_Subspace_Model}, we see that we can choose $K$ to be the natural ``cutoff" point (which is in general much smaller than $M$) that can be determined in a principled manner.  This gives us significant dimensionality reduction at almost no cost to the estimation error.

\begin{figure}[h]
        \centering
        \begin{subfigure}[b]{0.45\textwidth}
            \centering
            \includegraphics[width=\textwidth]{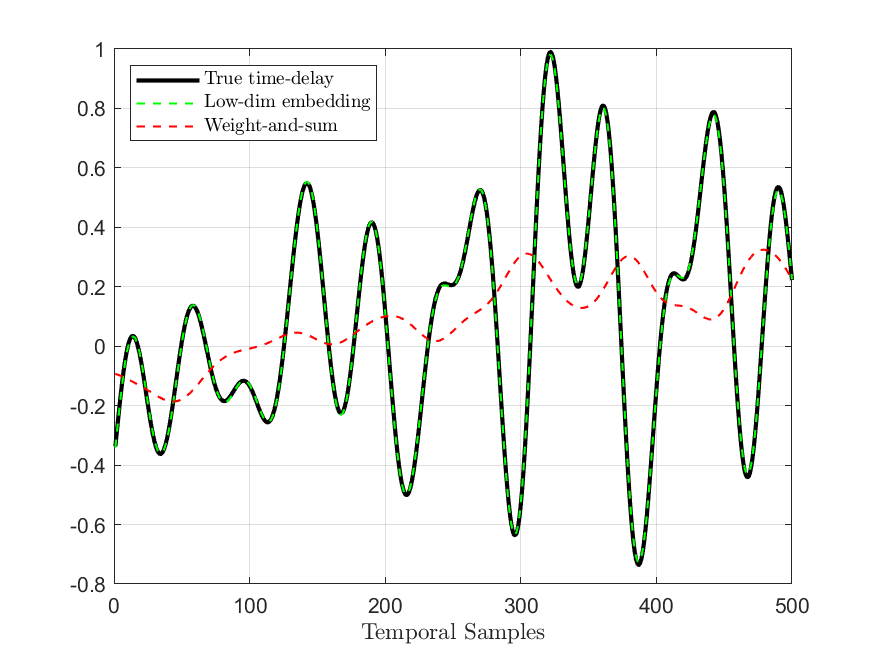}
            \caption%
            {{\small}}    
        \end{subfigure}
        \begin{subfigure}[b]{0.45\textwidth}  
            \centering 
            \includegraphics[width=\textwidth]{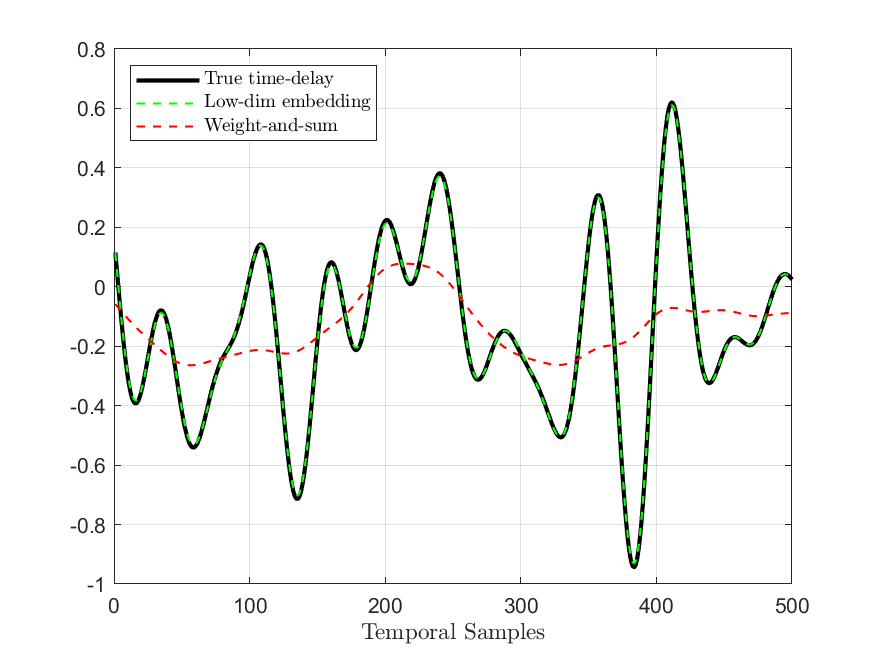}
            \caption%
            {{\small}}    
        \end{subfigure}

        \begin{subfigure}[b]{0.45\textwidth}
            \centering
            \includegraphics[width=\textwidth]{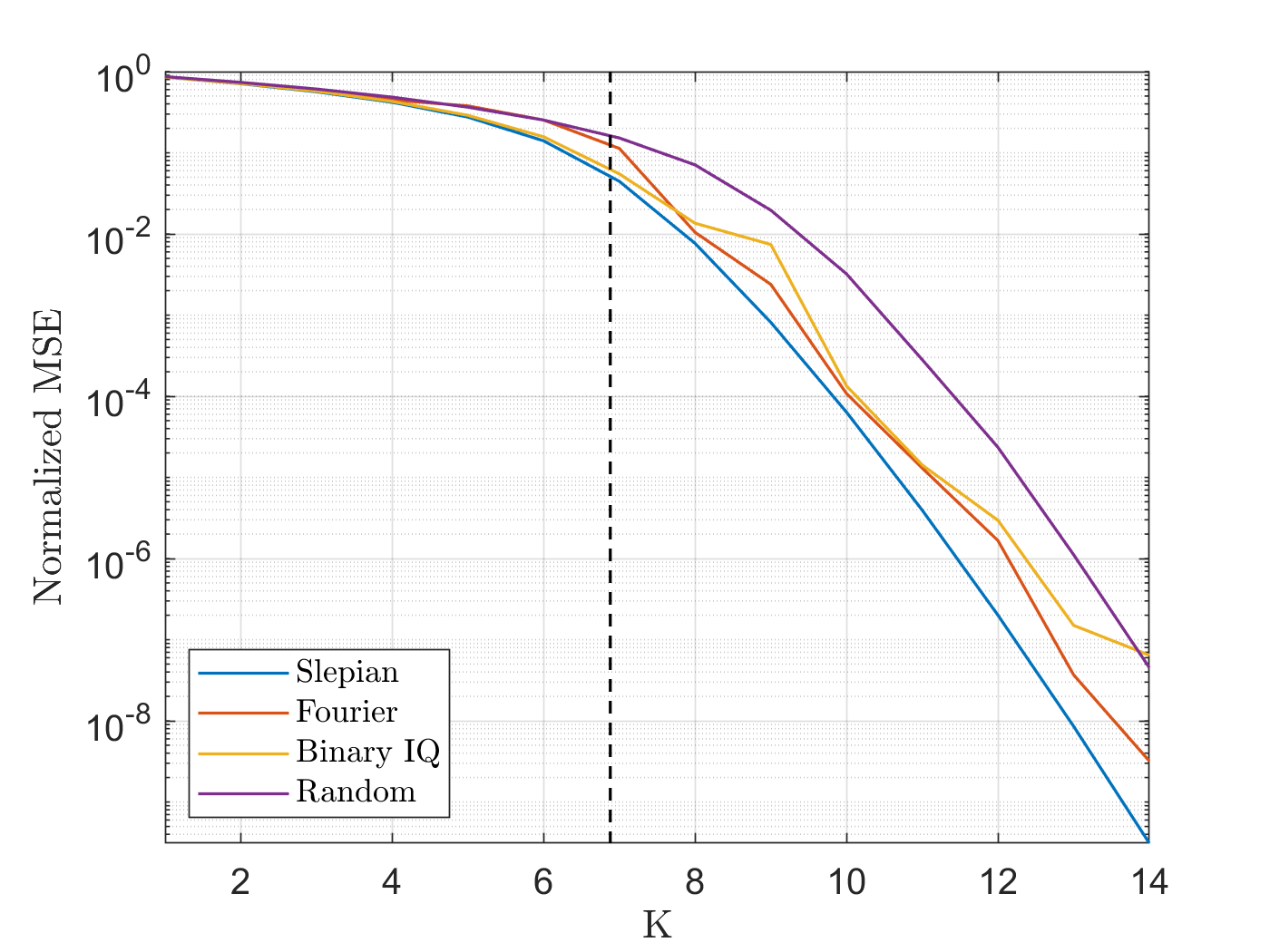}
            \caption%
            {{\small}}    
        \end{subfigure}
        \begin{subfigure}[b]{0.45\textwidth}
            \centering
            \includegraphics[width=\textwidth]{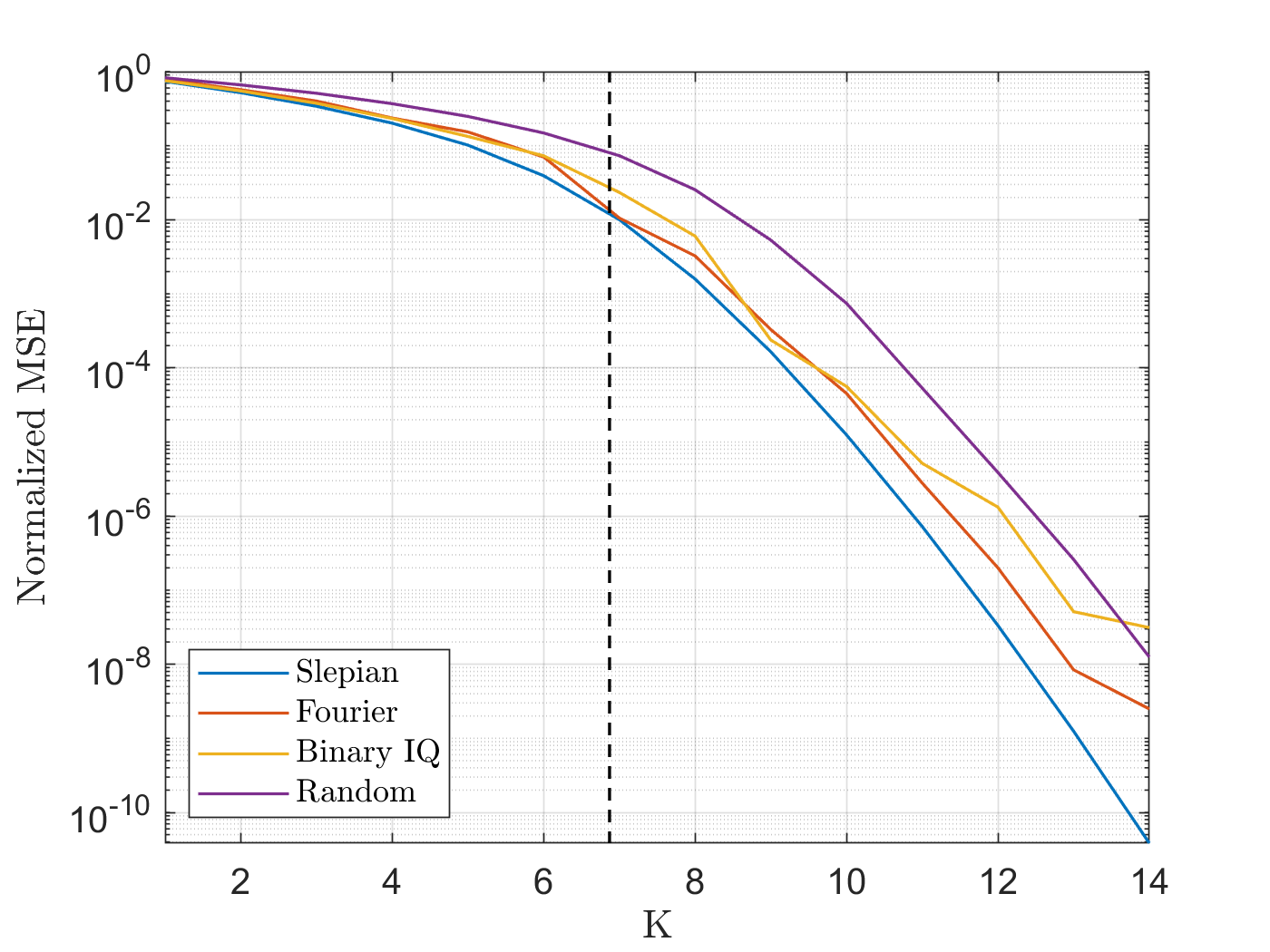}
            \caption%
            {{\small}}    
        \end{subfigure}
        \caption[Projection]
        {\small\sl  Qualitative example of signal beamformed via $K=7$ Slepian measurements compared to true time delay and weight-and-sum for (a) 64-element ULA and (b) $16\times 16$ UPA. For the same incident signals repectively the  MSE according to \eqref{eq:MSE} under a variety of $\mPhi$ with the time-bandwidth product indicated by the dashed black line for (c) 64-element ULA and (d) $16\times 16$ UPA. } 
        \label{fig:Recovery}
\end{figure}

As a concrete demonstration of the utility of \eqref{eq:MSE_sol} we consider two example cases: a 64 element ULA and a $16\times16$ UPA. For the ULA case we the incident signal has parameters $f_c=28$ Ghz, $\Omega=3$ GHz, and $\theta=0$ (i.e. broadside). Similarly for the UPA the incident signal has parameters $f_c=28$ Ghz, $\Omega=8.5$ GHz, and $\vtheta=[\pi/4,0]^T$. We have chosen the parameters in this manner such that in both case the sample snapshots coming off the array approximately lie in a subspace of dimension 7 in accordance with \eqref{eq:UPA_sub_dim} and \eqref{eq:ULA_sub_dim}. The noise in this example is considered to be negligible (e.g. $\sigma^2=0$) and we set $\mPhi$ to be the $K=7$ transposed dominate eigenvectors of $\mR$. A series of $\vy_0$ are estimated, after which the system performs digital beamforming via multiple fractional-delay filters to produce a single signal output. These results, shown for both cases in Figure~\ref{fig:Recovery}(a-b), is compared to a true-time delay and a narrowband beamforming implementation. The true-time delay represents the optimal in terms of coherent array processing, and as can be visually determined our proposed method yields an almost identical result. On the other hand the narrowband implementation, which may be naively implemented as an approximation to either method, produces a highly distorted signal.

While there is a clear choice for the best $\mPhi$ given the number of rows allowed, there may be constraints on how $\mPhi$ is implemented.  The expression \eqref{eq:MSE} gives us an unequivocal way to judge a candidate $\mPhi$ for a given $\mR$ (recall that $\mR$ depends on properties of the incoming signal, its direction of arrival, and the geometry of the array). To this end we examine three additional measurement types: unimodular, binary-IQ, and random. A unimodular $\mPhi$ has elements of the form $\mPhi[k,m] = e^{j2\pi f_k \tau_m}$ such that there is no tapering of the measurements. A further simplification of this comes in the form of binary-IQ or ``2-bit" measurements where $\mPhi[k,m] \in \{1,-1,j,-j\}$. A caveat of this form of measurement is that designing a $\mPhi$ to approximate the range of the Slepian basis vectors under the given constraints is a particularly difficult task, as will be discussed in further detail in Section~\ref{sec:Simplified_Measurements}. To produce random measurements we simply draw the components of $\mPhi$ i.i.d. from a complex Gaussian distribution.

To quantify the performance under a variety of measurement types we return to the two signals described in our previous example, again incident to a 64-element ULA and $16\times16$ UPA respectively. The signal statistics remain fixed while we vary the number of measurements $K$ for each choice in $\mPhi$. The MSE, as caclulated by \eqref{eq:MSE} and normalized by $\trace{(\mR)}$, is shown for both array configurations in Figure~\ref{fig:Recovery}(c-d). These results indicate that the alternative measurement schemes are able to achieve similar levels of distortion to the optimal Slepian measurements but at the cost of requiring more measurements. Hence we pay for our reduction in measurement complexity by requiring more measurements.

%--------------------------------------------------------------------------

%--------------------------------------------------------------------------
% SECTION: Temporal Decimation
\section{Temporal decimation}
\label{sec:Temporal_Decimation}

The Slepian embedding and the recovery framework discussed in Sections~\ref{sec:Slepian_Subspace_Model} and \ref{sec:Measurement_and_recon} above show how we can take advantage of spatial correlations in a single snapshot $\vy(t)$ to reduce array read out.  In this section, we discuss how the temporal relationship between the snapshots can be used to reduce the subsequent sampling rate.

\begin{figure}[h]
    \centering
    \includegraphics[width=\textwidth]{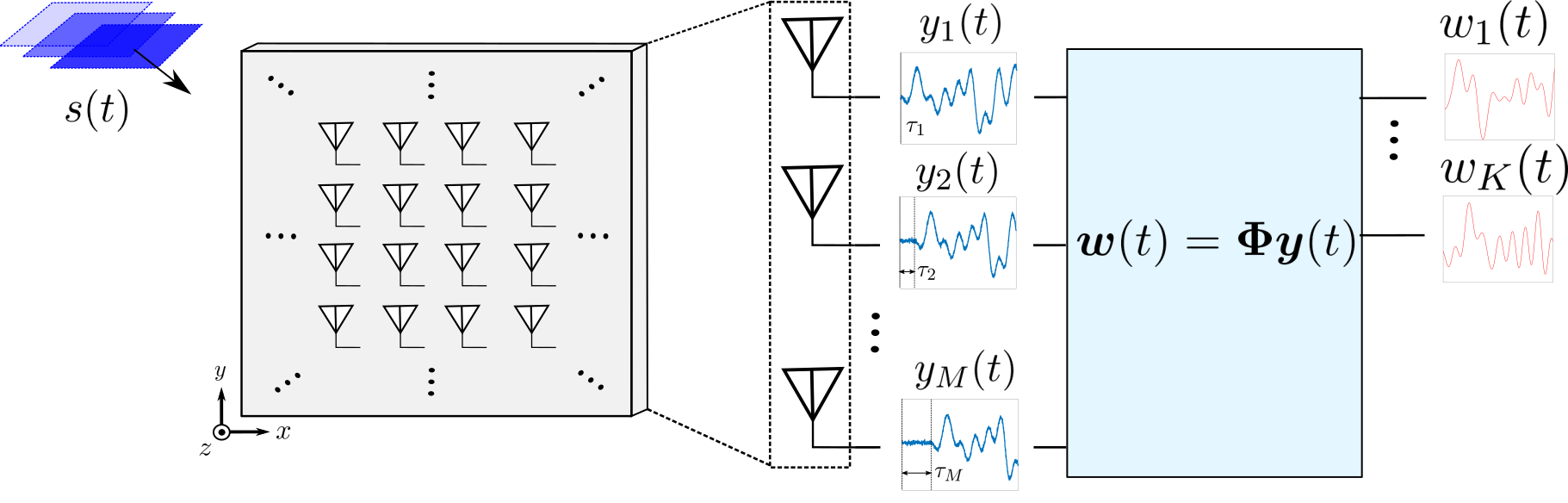}
    \caption{\small \sl The proposed measurement system acquires $K$ different weighted summations of the $M$ sensor outputs from the array where in general $K \ll M$. Though substantially smaller in dimension than the ambient array the output signal $\vw(t)$ captures nearly all the necessary information needed to reconstruct a full representation of $\vy(t)$.}
    \label{fig:LowDim_Beamformer}
\end{figure}

We have seen that with a single signal $s(t)$ incident on the array, a snapshot $\vy(t)$ consists of a set of $M$ (nonuniform, in general) samples of the signal over a time interval of length $A_{\vtheta}/c$, with $A_{\vtheta}$ given by \eqref{eq:effective_aperture}.  We have also seen that the relationships between the samples in a single snapshot allow us to compress them by projecting into a low dimensional subspace (as in \eqref{eq:Arbitruary_sub_dim}) and that we can recover them with essentially no loss.  If the linear encoding is performed with an analog vector-matrix multiply, we have transformed the $M$ signals $y_1(t),\ldots,y_M(t)$ coming off of the array elements into $K$ encoded signals $w_1(t),\ldots,w_K(t)$ as depicted in Figure~\ref{fig:LowDim_Beamformer}.  The bandwidth of the $w_k(t)$ is the same as that of the $y_m(t)$ (which is also the same as the incident signal), and taking samples of the $w_k(t)$ provides us with the same information as sampling the $y_m(t)$ directly.

\begin{figure}[h]
    \centering
    \begin{subfigure}[b]{0.32\textwidth}
    \includegraphics[width = \textwidth]{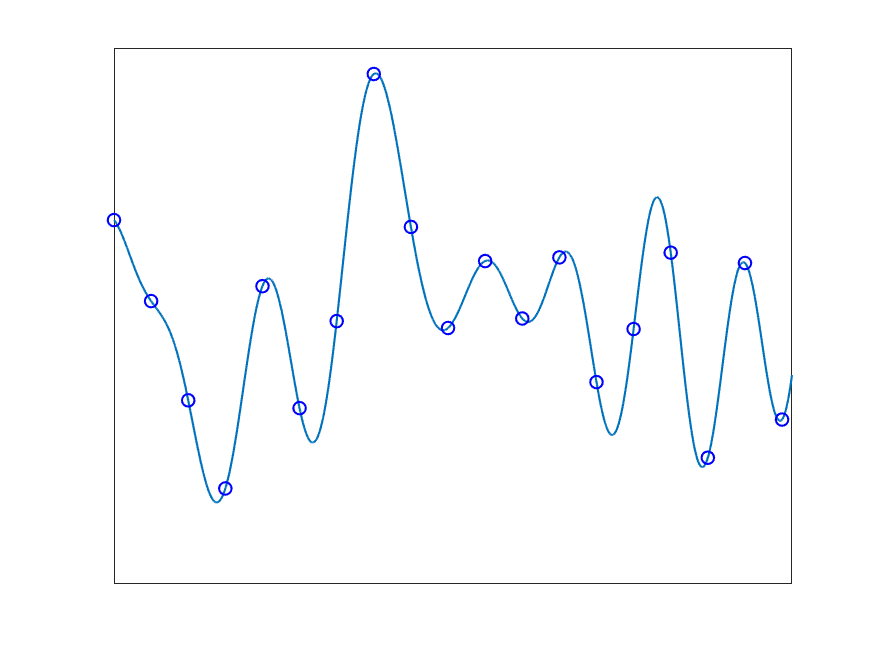}
    \caption{}
    \end{subfigure}
    \begin{subfigure}[b]{0.32\textwidth}
    \includegraphics[width = \textwidth]{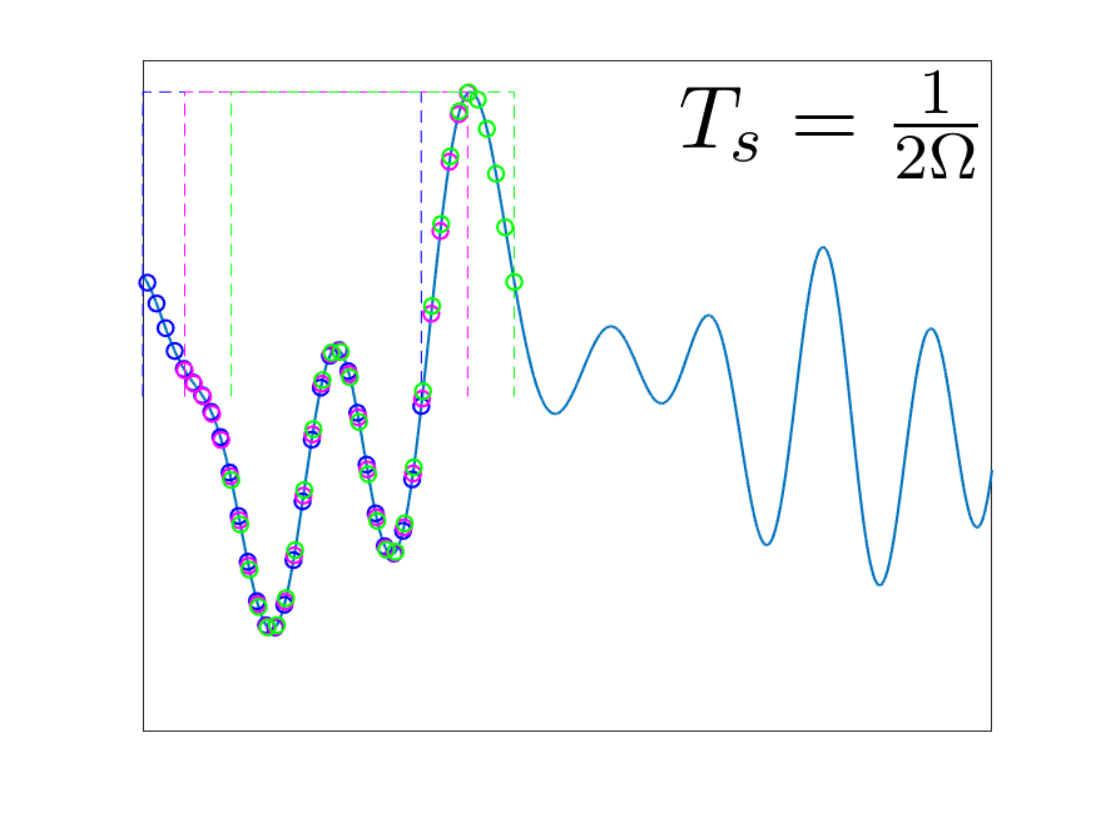}
    \caption{}
    \end{subfigure}
    \begin{subfigure}[b]{0.32\textwidth}
    \includegraphics[width = \textwidth]{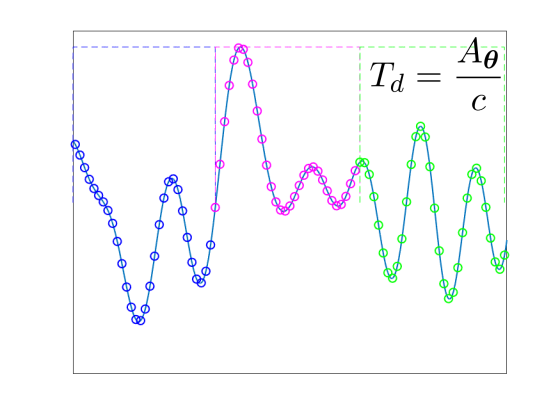}
    \caption{}
    \end{subfigure}
    \caption{ \small\sl(a) An example incident signal with bandwidth $\Omega$ with samples taken at the Nyquist rate overlayed.  (b) Temporal snapshots $\vy(t)$ taken by a 32-element array with $2\Omega A_{\vtheta}/c = 7$; the snapshots are taken at intervals of $T_s = \frac{1}{2\Omega}$.  (c) Temporal snapshots taken at the more relaxed rate of $T_d = A_{\vtheta}/c$.  We see that even a significant undersampling of the snapshots can yield enough samples to reconstruct the signal.
    }
    \label{fig:array_nyquist_dim}
\end{figure}

It is possible to capture the signal $s(t)$ by sampling the $w_k(t)$ at a rate far below the Nyquist rate.  Figure~\ref{fig:array_nyquist_dim} shows why this is true.  In Figure~\ref{fig:array_nyquist_dim}(a), we see an example of a signal with samples taken at the Nyquist rate.  In Figure~\ref{fig:array_nyquist_dim}(b), we see snapshots $\vy(t)$ taken in increments of $T_s = \frac{1}{2\Omega}$, corresponding to taking Nyquist rate samples of each of the $y_m(t)$ in parallel (which can again be reconstructed from the $w_k(t)$).  These snapshots overlap significantly, resulting in a signal that is heavily oversampled.  In Figure~\ref{fig:array_nyquist_dim}(c), we see that relaxing the sampling rate to $T_d = A_{\vtheta}/c$ so that the snapshots do not overlap still results in a sufficient number of samples to reconstruct the signal.  This can be interpreted as replacing the apparent loss of information from temporal subsampling with spatial samples.

Subsampling in time does have an obvious drawback: the very redundancy we are limiting is what is leveraged to give the classic array gain.  In reducing the sampling rate, we are losing some of our ability to coherently average out in-band noise.  However, it is often the case that a significant portion of the noise is caused by the ADC, and this noise becomes more pronounced at higher sampling rates \cite{Walden:1999}.  This presents a trade-off between the array gain and the quality (effective number of bits) of the samples that can be tuned depending on the particulars of the application.

%---
%{\color{red} Our previous discussion implicitly takes advantage of the spatial-temporal coupling of the array, where different spatial sampling points map to corresponding temporal instances of the underlying signal over an interval of width $\frac{A_e}{c}$. The correlation of samples within the snapshots $\vy(t)$ are ultimately what we leverage to produce our low dimensional encoding of the signal. We now transition from a discussion on intra-snapshot sample correlation to inter-snapshot sample correlation. Consider the signal shown in Figure \ref{fig:array_nyquist_dim}a overlayed with temporal samples taken at the Nyquist rate. We then compare this to three successive snapshots of signal taken from a 32 element ULA configured such that the underlying subspace dimension is 7. Figure \ref{fig:array_nyquist_dim}b shows that when the snapshots are acquired at the Nyquist rate the samples between snapshots highly overlay. If instead we acquire the snapshots at a decimated rate of $\frac{A_e}{c}$ the snapshots no longer overlay as depicted in Figure \ref{fig:array_nyquist_dim}c while still remaining highly correlated. This implies that by leveraging our limited array readout we can in turn reduce the temporal sampling rate substantially below the Nyquist rate.
%}

%-------
%-------

%%The principle is that because we are sufficiently sampling the signal over an interval with each snapshot, we do not need to produce a snapshot every at every Nyquist interval $T_s = \frac{1}{2 \Omega}$. We only really need to temporally sample at the decimated rate $T_d = \frac{A_e}{c}$ in order to accurately reconstruct the signal. Comparing this decimated rate to the Nyquist rate we see that
%\begin{align*}
%    \frac{T_d}{T_s} = \frac{2A_e\Omega}{c}
%\end{align*}
%which is exactly the number of linear combinations we take across the array to form the limited array readout. Hence, if we sample temporally at this decimated rate we supplement the apparent loss of information with spatial samples. The trade off is that we lose the ability to coherently average out in-band noise since the coherent averaging leverages the redundancy we have used to reduce the sampling rate. However, it is worth noting that a significant amount of the noise is due to ADC and this becomes more pronounced at higher sampling rates. By reducing the sampling rate by a factor proportional to the spatial subspace dimension we stand to drastically reduce the ADC sampling requirements. In turn we can use higher quality ADCs and increase the ENOB, allowing for more advanced backend processing schemes that may allow us to make up for this loss in array gain.
%}

\begin{figure}[h]
    \centering
    \begin{subfigure}[b]{0.32\textwidth}
    \includegraphics[width = \textwidth]{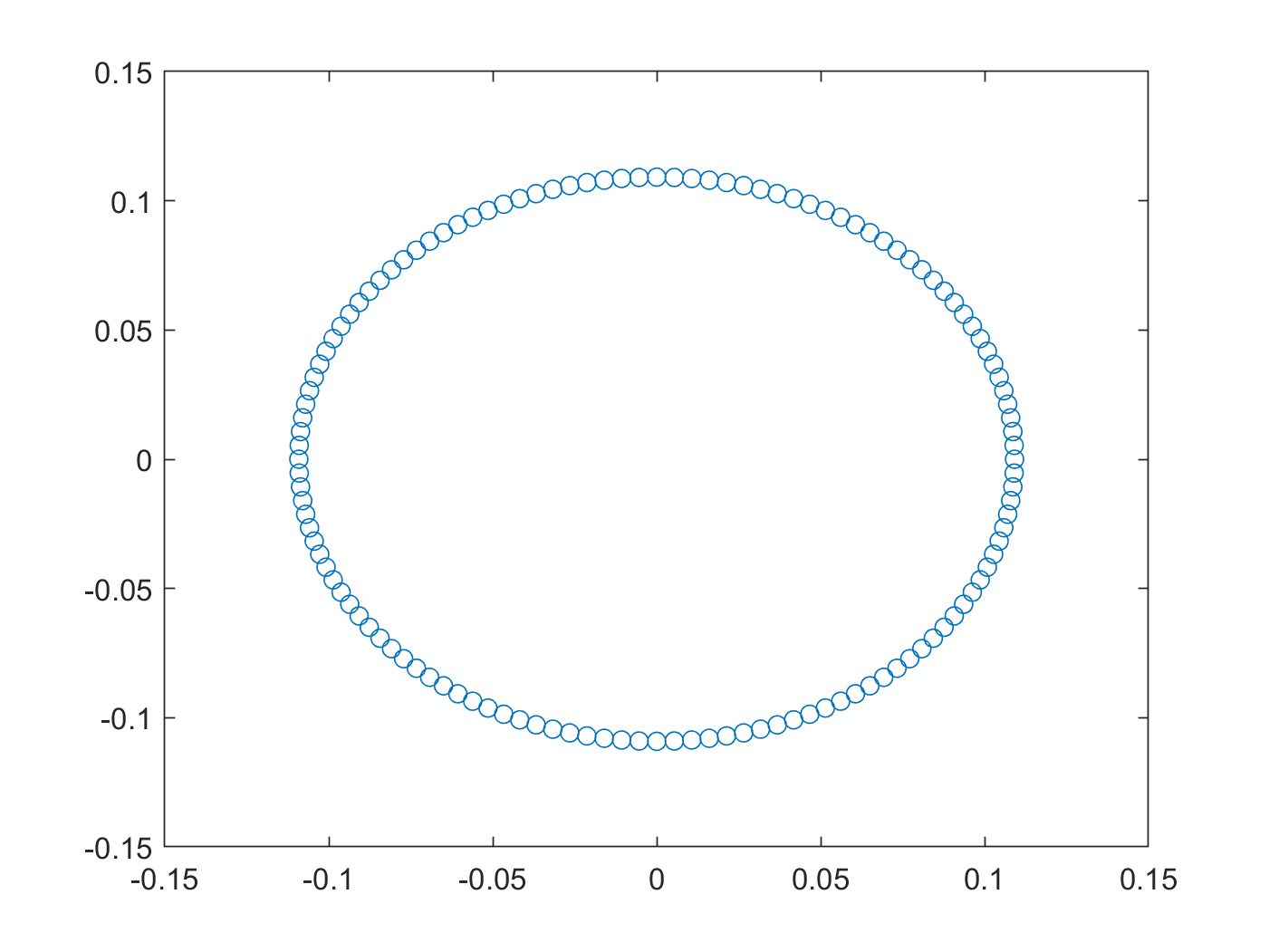}
    \caption{}
    \end{subfigure}
    \begin{subfigure}[b]{0.32\textwidth}
    \includegraphics[width = \textwidth]{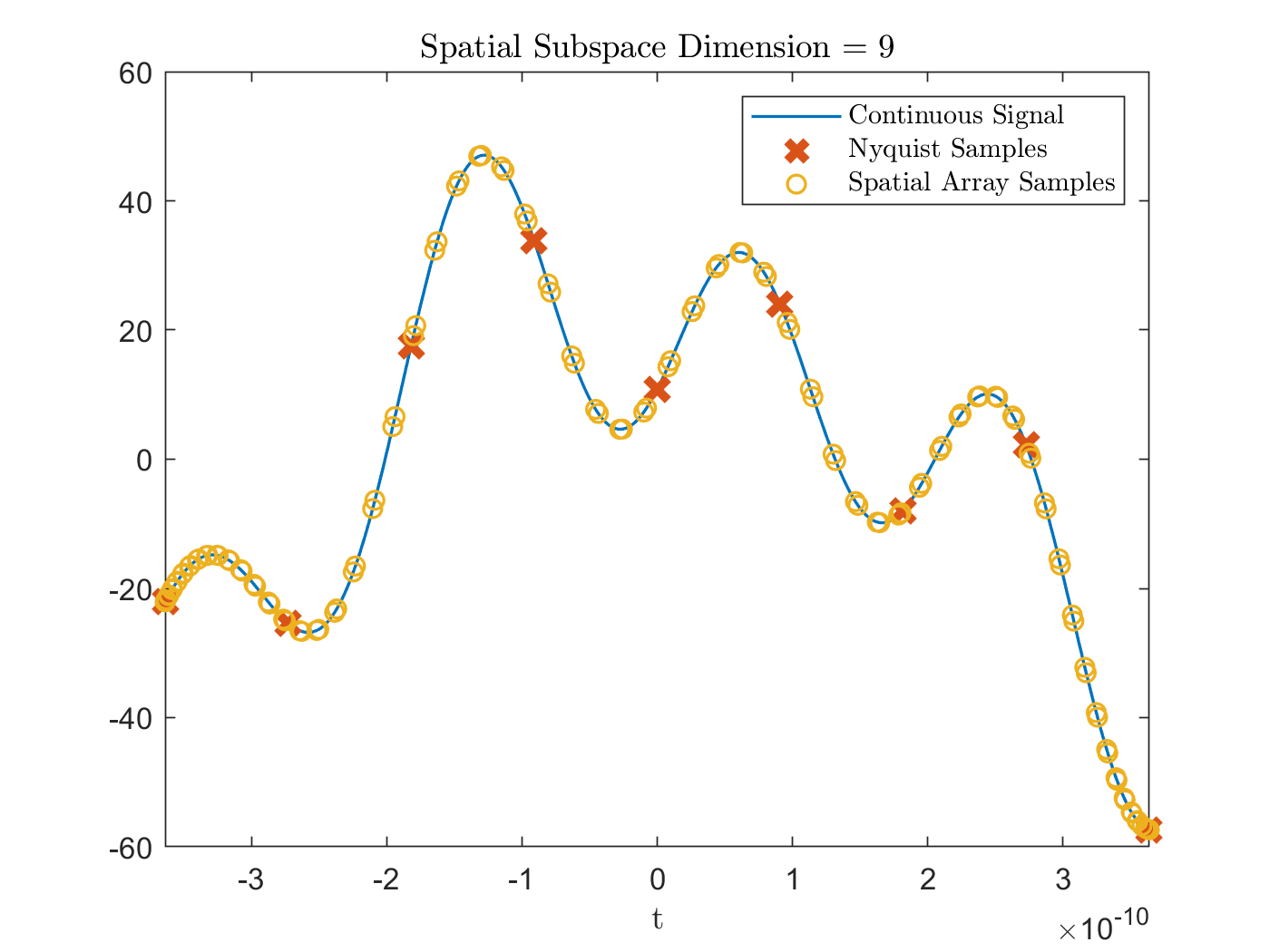}
    \caption{}
    \end{subfigure}
    \begin{subfigure}[b]{0.32\textwidth}
    \includegraphics[width = \textwidth]{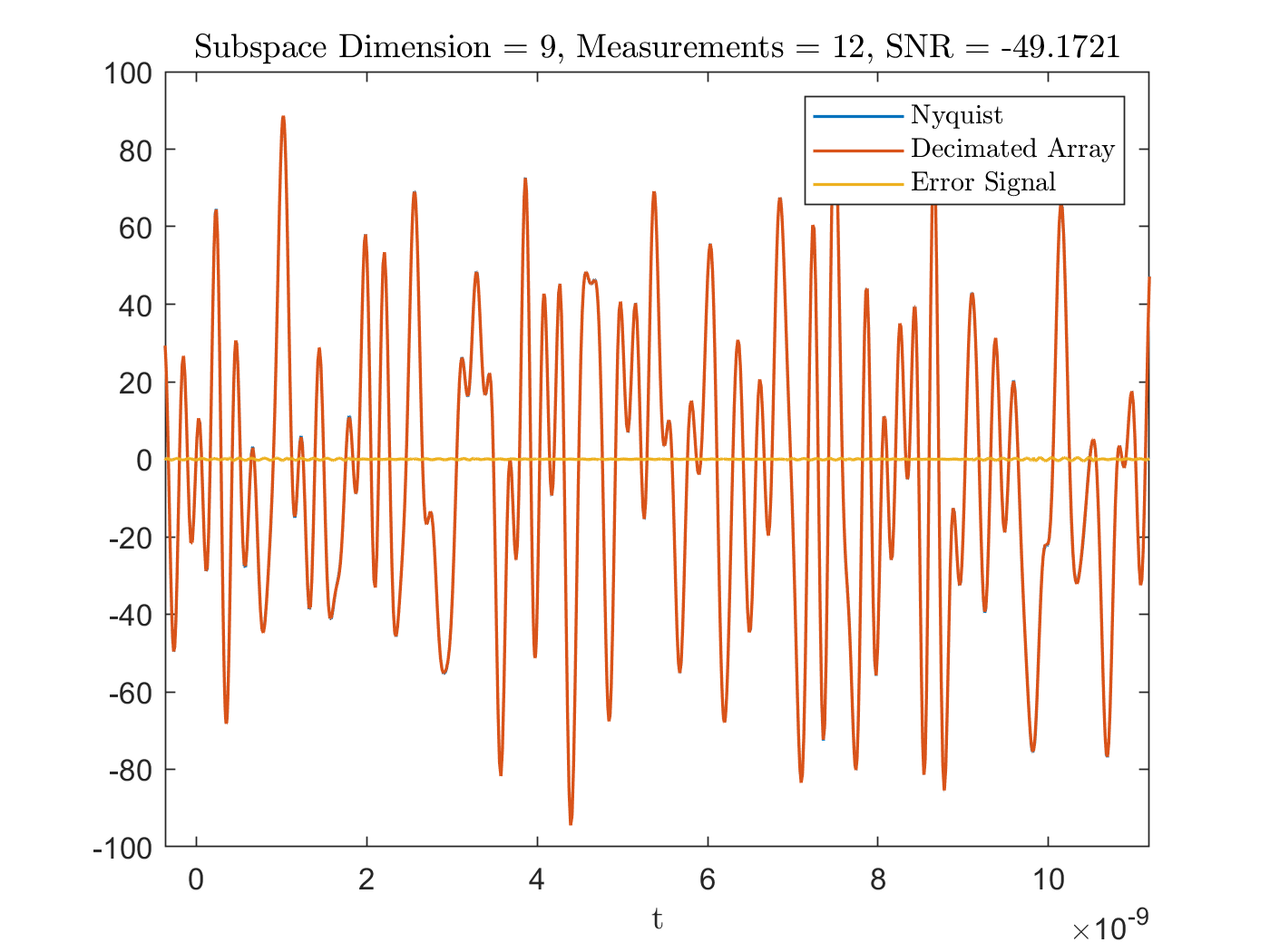}
    \caption{}
    \end{subfigure}
    \caption{\small \sl(a) Circular array geometry with standard $\frac{\lambda}{2}$ spacing. (b) Snapshot of the array compared to the Nyquist rate samples of the signal demonstrating the redundancy in the spatial samples. (c) Reconstructed signal compared to the Nyquist rate samples with the error overlayed.}
    \label{fig:conformal_array}
\end{figure}

As we have seen above in \eqref{eq:effective_aperture} and \eqref{eq:Arbitruary_sub_dim}, the dimension of the embedding subspace depends on the angle of incidence, as the interval of time $A_{\vtheta}/c$ over which the array sees the signal at any instant depends on $\vtheta$.  For linear and planar arrays, the variations in $A_{\vtheta}$ can be dramatic.  For the UPA, $A_{\vtheta}=0$ when the signal arrives broadside; in this case, we can achieve tremendous compression spatially (as $\operatorname{dim}(\setS_{\vtheta}) = 1$), but cannot reduce the sampling rate at all as every array element is observing the incoming signals at exactly the same place.  In general, the trade-offs available to us will depend on $\vtheta$, and taking advantage of these trade-offs might involve changing the sampling rate dynamically.  
There are, however, array architectures where this effect is lessened.  For example, the circular array shown in Figure~\ref{fig:conformal_array}(a) has the same effective aperture for all angles of arrival in the plane, making angle-independent temporal decimation possible.
%
%One can note that for planar and linear arrays the spatial subspace dimension is highly dependent on the angle of incidents. Signals coming in at broadside will always have a spatial subspace dimension of one while for signals coming from broadside the dimension is far higher. Therefore in order to leverage this decimated sampling scheme for these arrays we would require a variable sampling rate ADC, and at broadside we would require critical sampling. However, if we consider utilizing a conformal array there are several simple architectures that do not fall prey to this complication. 

Figure~\ref{fig:conformal_array} illustrates how temporal decimation might work.  
In Figure~\ref{fig:conformal_array}(a) illustrate a circular array consisting of $125$ elements placed at a $\frac{\lambda}{2}$ spacing for $f_c = 28$ GHz.  Suppose that the bandwidth of the incoming signal is $5.5$ GHz, meaning we would generally have to sample at $11$ GHz to acquire Nyquist rate samples. Leveraging the redundancy shown in Figure \ref{fig:conformal_array}(b), we reduce sampling rate by a factor roughly equal to the spatial dimension such that $f_s = 1.38$ GHz. Of course we can still embed each snapshot in a Slepian subspace as previously described to produce a limited array readout and further reduce complexity. We can then simultaneously decode the snapshots and interpolate the non-uniform samples therein onto the uniformly spaced Nyquist rate samples. The results of this reconstruction are shown in Figure \ref{fig:conformal_array}(c), and there is very little distortion compared to the signal sampled directly at the Nyquist rate. 
%Hence we can accurately represent the signal even when sampling at this reduced rate. 
As mentioned above, our ability to sample at this reduced rate is invariant under the angle of incidence due to the geometry of the array.

%-------
%-------

Finally, we note that there is a methodology for reconstructing the signal from overlapping blocks of non-uniform samples.  We start by breaking the signal apart over intervals of time $[t^\ell_i,t^r_i]$, with $t^\ell_{i} < t^r_{i-1} < t^\ell_{i+1}$ for all $i\in\Z$.  Inside of each frame, we represent the local piece of the signal using $N$ basis vectors $\{\psi_{i,n},~n=1,\ldots,N\}$, and so
\[
    s(t) = \sum_{i}\sum_{n=1}^N\alpha_{i,n}\psi_{i,n}(t).
\]
If we denote by $\vz_i$ all of the samples that lie in the interval $[t^r_{i-1},t^\ell_{i+1}]$, then $\vz_i$ give us information about the expansion coefficients $\valpha_{i-1} = \{\alpha_{i-1,n}\}_{n=1}^N$ and $\valpha_{i}$.  After observing samples through the $I$th interval, we can estimate the expansion coefficients $\valpha_i$ by solving the least-squares problem
\begin{equation}
    \label{eq:streamingls}
    \minimize_{\{\valpha_i\}}~\sum_{i=1}^I\|\vz_i - \mA_i\valpha_i - \mB_i\valpha_{i-1}\|_2^2,
\end{equation}
where $\mA_i$ and $\mB_i$ are $M\times N$ matrices with entries
\[
    \mA_i[m,n] = \psi_{i,n}(\tau_{i,m}), \quad 
    \mB_i[m,n] = \psi_{i-1,n}(\tau_{i,m}),
\]
where $\tau_{i,m}$ is the location of the $m$th sample in $\vz_i$.  The optimization program \eqref{eq:streamingls} can be solved in an online manner with matrix operations akin to the Kalman filter \cite{Hamam:2021}.  The framework can also be adjusted to work directly from compressed samples $\vw_i = \mPhi\vz_i$ by combining ideas from Section~\ref{sec:Measurement_and_recon} with those from \cite{Hamam:2021}.

% {\color{red}\bf [We might want to add a figure to show how the samples are divided up above.]}

%--------------------------------------------------------------------------
% SECTION: Fast Slepian Computations

\section{Fast Slepian computations}
\label{sec:Fast_Slepian_Computations}

The proposed method of broadband beamforming can be viewed as a two step process: measurement and reconstruction. The reconstruction process described in Section~\ref{sec:Measurement_and_recon} above, an in particular by \eqref{eq:MSE_sol}, involves a sequence of matrix operations.  These matrices can all be pre-computed, and so mapping the measurements $\vw$ to the sample estimates $\hat\vy$ requires the application of a known $M\times K$ matrix at a computational cost of $O(MK)$.  In this section, we discuss how recent innovations in Slepian computations can reduce this computational cost.

%We first embed the array snapshots in a low dimensional subspace to produce a limited array readout. Then we reconstruct the original snapshot such that we are left with a collection of samples from the incident signal. At this point we are free to apply numerous DSP techniques depending on the application. As will be shown, recent innovations in fast Slepian computations allow these operations to be performed in a computationally efficient manner.

{\bf Uniform linear arrays.}
We have seen, as illustrated in Figure~\ref{fig:prolate_eigenvalue_non_uniform}, that the covariance matrix $\mR$ in the reconstruction equation \eqref{eq:MSE_sol} is effectively rank deficient.  Choosing $K$ large enough, we can approximate $\mR$ as
\begin{equation}
    \label{eq:Rapprox}
    \mR \approx \mC_K\mU_K^\H,
\end{equation}
where $\mC_K$ and $\mU_K$ are $M\times K$ matrices.  In fact, the recent works \cite{karnik2017fast,zhu18ro} show that we can take
\begin{equation}
    \label{eq:CK}
    \mC_K = \begin{bmatrix} \mF_{M,W} & \mL \end{bmatrix},
\end{equation}
and $\mU_K$ of a similar form, where $\mF_{M,W}$ is the first $\ceil{2\Omega A_{\vtheta}/c}$ columns of the standard $M\times N$ discrete Fourier transform (DFT) matrix, and $\mL$ is a known $M\times \mathcal{O}(\log M \log \frac{1}{\epsilon})$ matrix and have $\|\mR-\mC_K\mC_K^\H\|\leq\epsilon$.  The key thing to recognize here is that (after fixing $\epsilon$ to be something sufficiently small) $\mC_K$ can be applied in $O(M\log M)$ time: $\mF_{M,W}$ with a (partial) fast Fourier transform and the ``skinny'' $M\times O(\log M)$ matrix $\mL$ using a standard vector-matrix multiply.  We can thus approximate \eqref{eq:MSE_sol} as
\[
    \hat\vy \approx \mC_K\left(\mU_K^\H\mPhi^\H(\mPhi\mR\mPhi^\H + \sigma^2\mId)^{-1}\right)\vw
\]
by applying the known $K\times K$ matrix $\mU_K^\H\mPhi^\H(\mPhi\mR\mPhi^\H + \sigma^2\mId)^{-1}$ to $\vw$ then using a fast application of $\mC_K$ for a total cost of $O(M\log M + K^2)$.  For $\log M < K \ll M$, the approximation given by \eqref{eq:Rapprox} and \eqref{eq:CK} can result is significant computational savings.

{\bf General array geometries.} Though not explicitly stated in the recent works on fast Slepain computations, fast computations are also possible for general arrays that produce non-uniform samples of the signal.  As previously mentioned, the discrete time spectral concentration property converges to its continuous time analog as the sampling grid becomes arbitrarily fine \cite{karnik2020improved,Bonami:2021,Boulsane:2019}.  This means that a bandlimited signal observed on the continuum over a finite segment of time can be represented using a small number of complex sinusoids (Fourier series coefficients) and a small number of smooth auxiliary functions.  An arbitrary set of samples of the bandlimited signal can be closely approximated by the corresponding samples of the sinusoids and the auxiliary functions.  This means that the covariance $\mR$ can be approximated as in \eqref{eq:Rapprox}, but with
\[
    \mC_K = \begin{bmatrix} \tilde\mF_{M,W} & \tilde\mL \end{bmatrix},
\]
where $\tilde\mF_{M,W}$ is a non-uniform (partial) DFT matrix.  There are fast non-uniform FFT methods for applying $\tilde\mF_{M,W}$ (see \cite{Keiner:2009} for a state-of-the-art implementation), and so $\hat\vy$ can again be estimated in $O(M\log M + K^2)$ time.

%In turn, this implies that the DFT operation and low-rank correction has an analog on the continuum as a Fourier series and some small number of correction functions. When the samples transition to a non-uniform placement we are simply sampling these continuous functions at different points. Therefore, with a few caveats the methodology used to derive fast techniques in the uniform case can be applied to deriving non-uniform analogs. One such caveat is that $\mF_{M,W}$ now gets replaced with a non-uniform DFT matrix $\tilde\mF_{M,W}$. However, there are existing algorithms for computing non-uniform DFTs in a comparable time to the standard FFT \cite{Keiner:2009}. Another caveat is that it is unclear how to produce a closed form expression for $\mL_1,\mL_2$ in the non-uniform case. Though they can still be pre-computed, the loss in structure makes it difficult to generalize their construction.

{\bf Non-uniform to uniform samples.}  The computational techniques above give us a fast way to recover the (in general nonuniform) samples corresponding to a single array snapshot from their linear measurements.  As most back-end signal processing operations are expecting uniform samples, we would like a fast method for mapping the non-uniform samples $\hat\vy$ to the associated uniform samples $\hat\vy_u$.  We discussed how this might be set up as a streaming problem in Section~\ref{sec:Temporal_Decimation} above, but there has also been recent work on how this problem can be solved in an efficient manner for a fixed batch of non-uniform samples.

A computationally efficient method for producing $\hat\vy_u$ from $\hat\vy$ via conjugate gradient descent (CGD) is described in \cite{Karnik:2019:nonuniform}. In particular, leveraging the fact $\hat\vy$ lies in a low dimensional Slepian space the authors showed that given a regularization parameter $\delta$ CGD converges in $\mathcal{O}\big (\text{polylog} \big( \frac{2\Omega A_{\vtheta}}{c}, \frac{Mc}{2\Omega A_{\vtheta}}, \frac{1}{\delta},\allowbreak \frac{1}{\epsilon}\big) \big)$ iterations to an error bounded by $\epsilon\norm{\hat\vy}_2$. Furthermore each iteration of CGD can be computed implicitly in $\mathcal{O}(M \log\frac{1}{\zeta})$ where $\zeta$ is an approximation error term. Thus we can efficiently produce uniform samples from our non-uniform snapshots.

\begin{figure}[h]
    \centering
    \begin{subfigure}[b]{0.48\textwidth}
    \includegraphics[width = \textwidth]{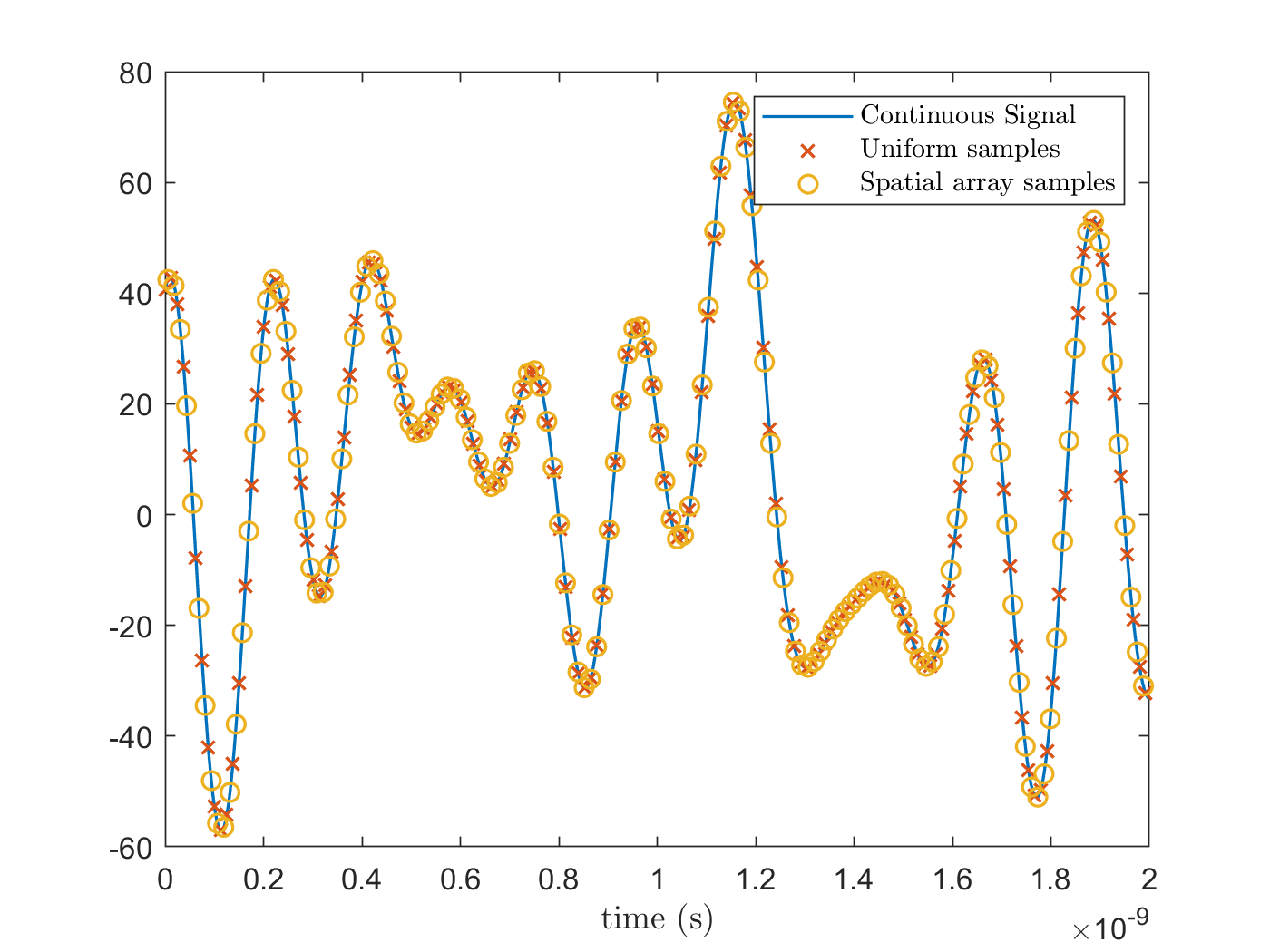}
    \caption{}
    \end{subfigure}
    \begin{subfigure}[b]{0.48\textwidth}
    \includegraphics[width = \textwidth]{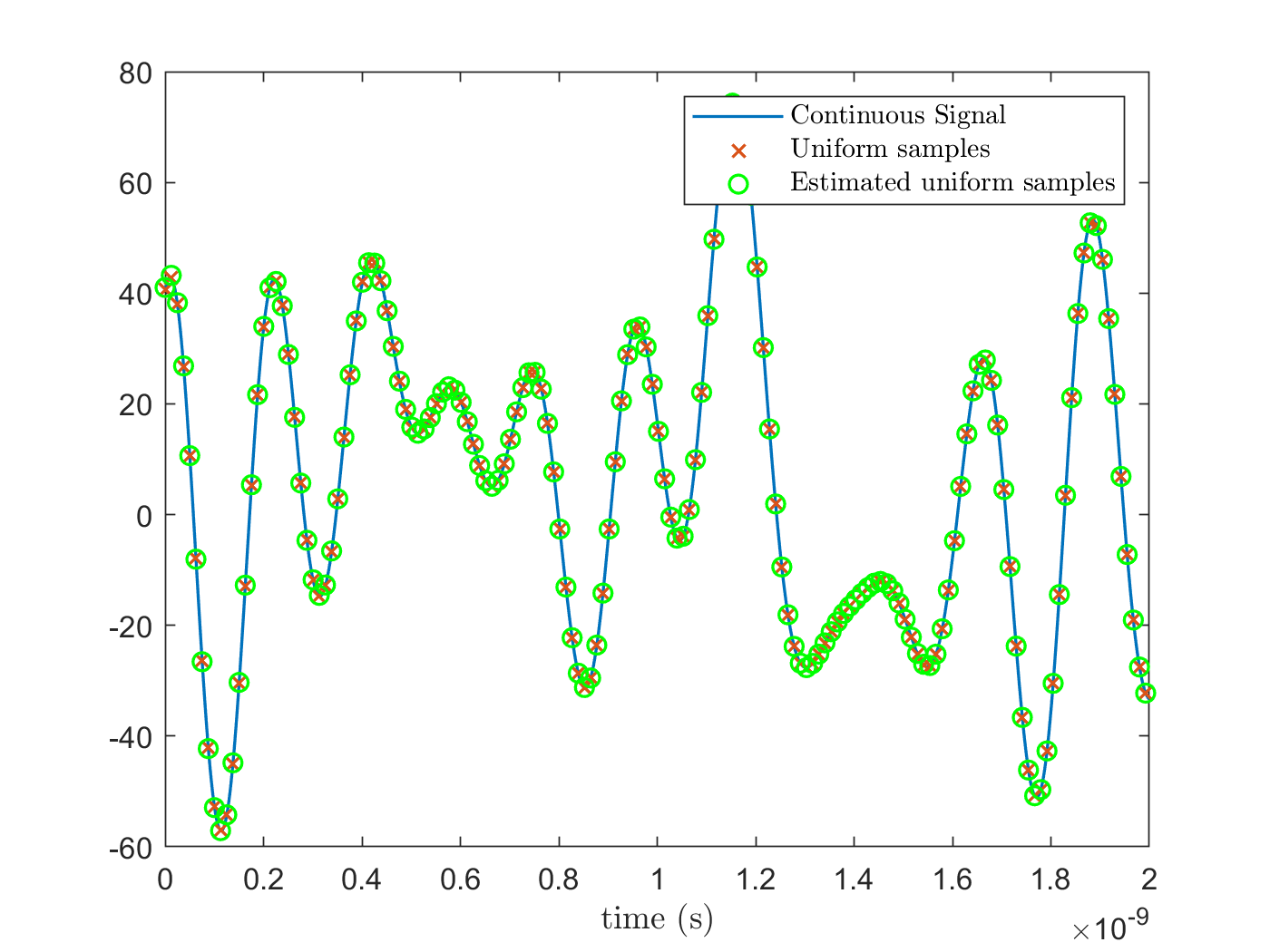}
    \caption{}
    \end{subfigure}
    \caption{\small \sl (a) A uniform grid of samples compared to the non-uniform samples reconstructed from a limited array readout. (b) True uniform samples compared to estimated samples recovered via the fast CGD descent algorithm.}
    \label{fig:Fast_reconstrunction_test}
\end{figure}

As a demonstration we examine a 5.5 GHz signal residing at a center frequency of 28 GHz incident to a $32 \times 32$ UPA. Each snapshot of the signal approximately lies in a 9-dimensional Slepian space, and we temporally sample at $1.28$ GHz meaning the signal is drastically under sampled in a traditional sense. We first embed each snapshot using $\mPhi = \mV_K^H$ and recover using \eqref{eq:MSE_sol}. We invoke the fast recovery algorithm to produce the uniform samples from each reconstructed snapshot. A window of spatial array non-uniform samples with a corresponding uniform grid of samples is shown in Figure \ref{fig:Fast_reconstrunction_test}(a). The recovered uniform samples are shown in Figure \ref{fig:Fast_reconstrunction_test}(b) and have a recovery SNR of 60 dB while CGD converged in a mere 26 iterations. Hence we were able to quickly resolve the uniform samples to within a modest error while traditionally under sampling the data. 

%-----
%-----

%% jrom: we covered this sufficiently in the previous section
%This is by no means represents an optimal methodology for producing a uniform sample output from multiple snapshots. For instance we have treated each snapshot independently, and hence neglected any correlation between successive snapshots. Explicitly taking into account the interplay between snapshots we can formulate this problem efficiently in a ``streaming reconstruction" manner \cite{Hamam:2021}. Under this framework each newly obtained snapshot is used to not only form a new estimate but also update all previous estimates, which ultimately yields a more accurate result.

%--------------------------------------------------------------------------
% SECTION: Signal Isolation

\section{Signal isolation}
\label{sec:Signal_Isolation}

A common method of quantifying a beamformer's performance is examining its ability to spatially isolate signals. Often termed ``sidelobe" behavior, this gives us a notion of how well the system is able to block out signals that are coming from directions outside the region of interest whether it be noise or an interferer. It is our goal in this section to formally characterize the behavior of our proposed beamformer in the presence of interferering signals with respect to the design of $\mPhi$. 

We start with a modification to our previous measurement model in which the observations combine samples from $L$ different signals,  
\begin{align}
    \label{eq:sig_interferer_model}
    \vw = \mPhi\left(\sum_{\ell=0}^{L-1} \vy_\ell\right) + \veta,
\end{align}
where $\vy_0$ corresponds to the array snapshot of  our signal of interest and $\{\vy_\ell\}_{\ell=1}^{L-1}$ are snapshots of undesired interfering broadband signals coming from distinct directions of arrival. Each $\vy_\ell$ is comprised of samples from independent complex gaussian random processes such that $\vy_\ell\sim \mathcal{N}_{\mathcal{C}}(\mtx{0},\mR_{\ell})$. Assuming the associated power spectral densities are flat then $\mR_{\ell} = \gamma_\ell \mE_{\ell} \mB_{\ell}\mE_{\ell}^H$ where $\mB_{\ell}$ is the prolate matrix associated with an angular bandwidth of $W_{\ell}$. Similarly $\mE_\ell$ is a diagonal modulation matrix associated with the angular center frequency $f_\ell$\footnote{We emphasize that $f_\ell$ and $W_\ell$ implicitly depend on angle as described in Section~\ref{sec:Slepian_Subspace_Model} even though this dependence is not explicit in notation.}. 
The parameter $\gamma_\ell$ represents the power of the signal. We note that in general $f_\ell$, $W_\ell$, $\gamma_\ell$, and DOA of each component signal are different. Due to the independence of the underlying processes the covarience matrix of this composition of signals is simply
\begin{align*}
    \mR  = \gamma_0\mE_{0} \mB_{W_0}\mE_{0}^H + \sum_{\ell = 1}^{L-1} \gamma_\ell \mE_{\ell} \mB_{\ell}\mE_{\ell}^H = \mR_0 + \mR_{I}.
\end{align*}
Returning to the estimation problem outlined in Section~\ref{sec:Measurement_and_recon}, with this new signal model in hand the optimal estimate of $\hat\vy_0$ (treating the other signals as part of the noise) becomes
\begin{align*}
    \hat\vy_0 = \mR_0 \mPhi^H(\mPhi\mR_0\mPhi^H + \mPhi\mR_I\mPhi^H + \sigma^2 \mtx{I})^{-1} \vw
\end{align*}
while associated MSE is
\begin{align}
    \label{eq:interferer_MSE}
    \mathbb{E}_{\vy_0,\dots,\vy_{L-1},\veta}\left\{\norm{\hat\vy-\vy_0}_2^2 \big | \vw \right\} = \trace{(\mR_0-\mR_0 \mPhi^H(\mPhi\mR_0\mPhi^H + \mPhi\mR_I\mPhi^H + \sigma^2 \mtx{I})^{-1}\mPhi\mR_0)}.
\end{align}

Interestingly, if we further assume that the array has been ``focused" on $\vy_0$ such that $\mPhi = \mV_K^H$ then the additional $\mPhi\mR_{I}\mPhi^H$ term in \eqref{eq:interferer_MSE} will only have a mild affect on the MSE in most cases. A detailed discussion on this is provided in Appendix~\ref{apx:mse_large_K} and argues this point based on the spectral properties of $\mR_{I}$ and $\mR_0$. In essence, when the signal and interferer bands are well separated in angular space\footnote{By separated in angular space we mean $|f_{\ell}-f_{\ell'}|\geq W_{\ell}+W_{\ell'}$ for all $\ell \neq \ell'$} for any particular column $\vv_k$ of $\mV_K$ we expect $\vv_k^H\mR_0\vv_k = \gamma_0\lambda_k$ to be large only when $\vv_k^H\mR_{I}\vv_k$ is comparatively small in magnitude. From this relationship we can conclude that the interferer more drastically affects the reconstruction error along the lower variance principle axis of $\mR_0$, which in turn have a smaller affect on the overall MSE by assumption. However, the degree to which the MSE is increased by the interferer ultimately depends on the dynamic range\footnote{Dynamic range meaning the relative magnitude of $\gamma_0$ compared to $\gamma_\ell$ for $\ell>0$}.

% To make this more explicit, consider a ULA and the ``worst case" bound $\mR_I\preceq \gamma_{max}(\mtx{I}-\gamma_0^{-1}\mtx{R}_0)$, then the MSE can be bounded by
% \begin{align}
%     \label{eq:interferer_MSE_bound_ULA}
%     \mathbb{E}_{\vy_0,\dots,\vy_{L-1},\veta}\left\{\norm{\hat\vy-\vy_0}_2^2 \big | \vw \right\} \leq 
%     \sum_{m=1}^K \gamma_0 \lambda_m \left ( 1 - \frac{\gamma_0\lambda_m}{\gamma_0\lambda_m + \gamma_{max}(1-\lambda_m) + \sigma^2} \right) + \sum_{m=K+1}^M \gamma_0 \lambda_m.
% \end{align}
% Due to the spectral concentration of $\mR_0$ we can assume that $1-\lambda_m$ is exceptionally small for $m<\ceil{2MW_{\theta_0}}$ and will increase to a significant quantity as we increase the index past this point. Therefore only the smaller eigenvalues in the sum will be appreciably damped, though this will largely depend on the dynamic range\footnote{Dynamic range meaning the relative magnitude of $\gamma_0$ compared to $\gamma_\ell$ for $\ell>0$}.

\begin{figure}[t]
        \centering
        \begin{subfigure}[b]{0.45\textwidth}  
            \centering 
            \includegraphics[width=\textwidth]{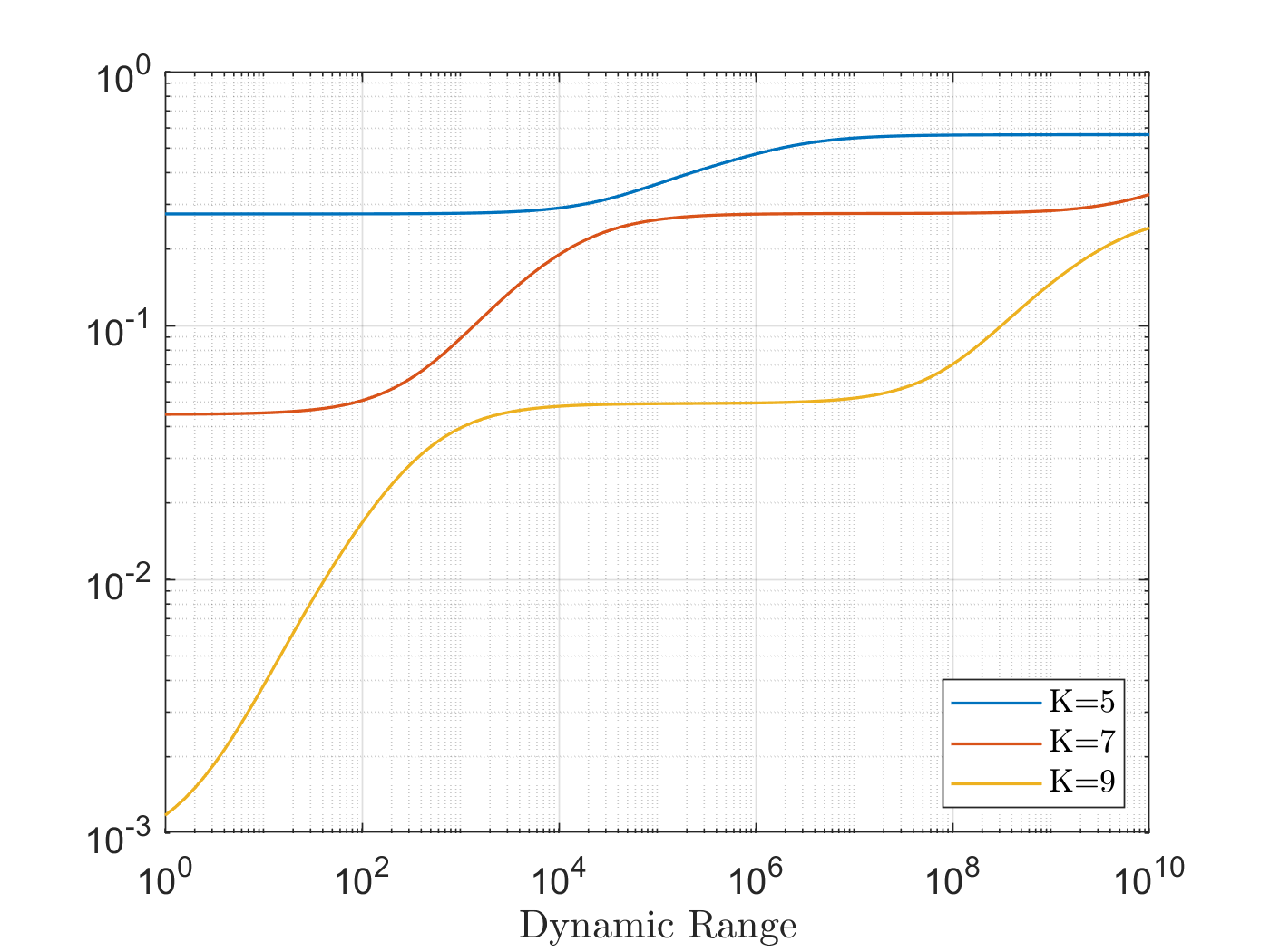}
            \caption%
            {{\small}}    
        \end{subfigure}
        \begin{subfigure}[b]{0.45\textwidth}  
            \centering 
          \includegraphics[width=\textwidth]{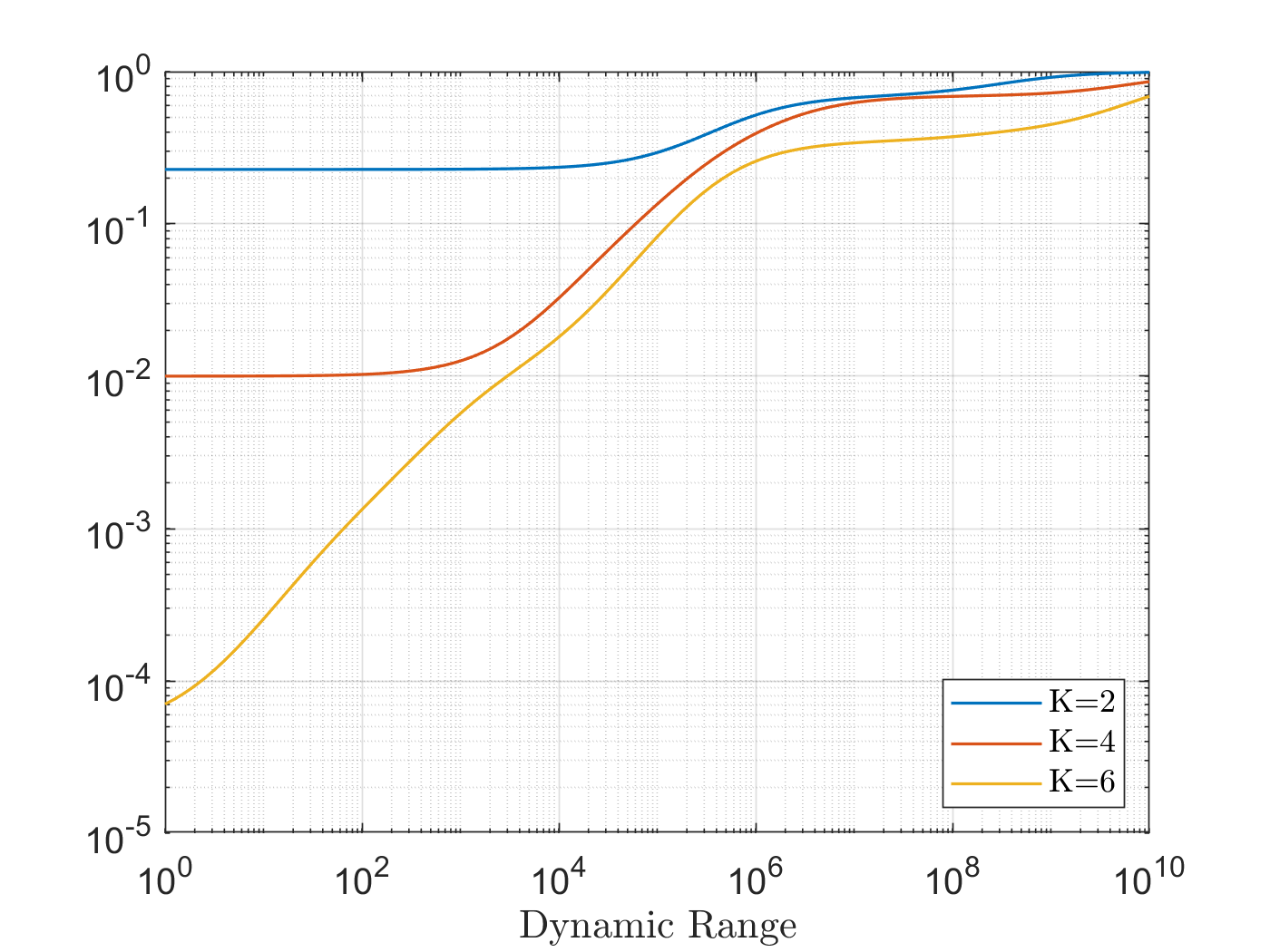}
            \caption%
            {{\small}}    
        \end{subfigure}
        
        \begin{subfigure}[b]{0.45\textwidth}  
            \centering 
            \includegraphics[width=\textwidth]{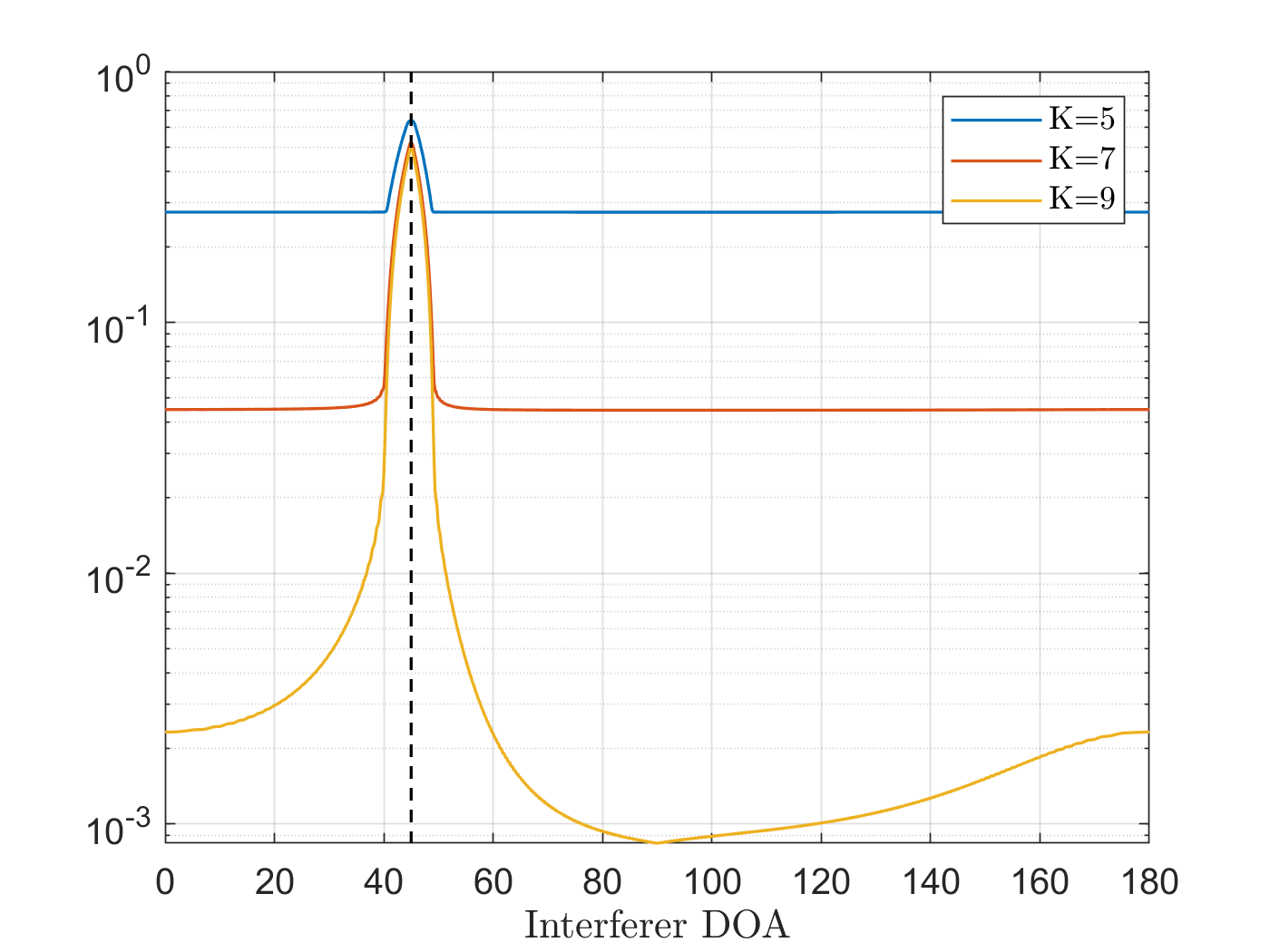}
            \caption%
            {{\small}}    
        \end{subfigure}
        \begin{subfigure}[b]{0.45\textwidth}  
            \centering 
           \includegraphics[width=\textwidth]{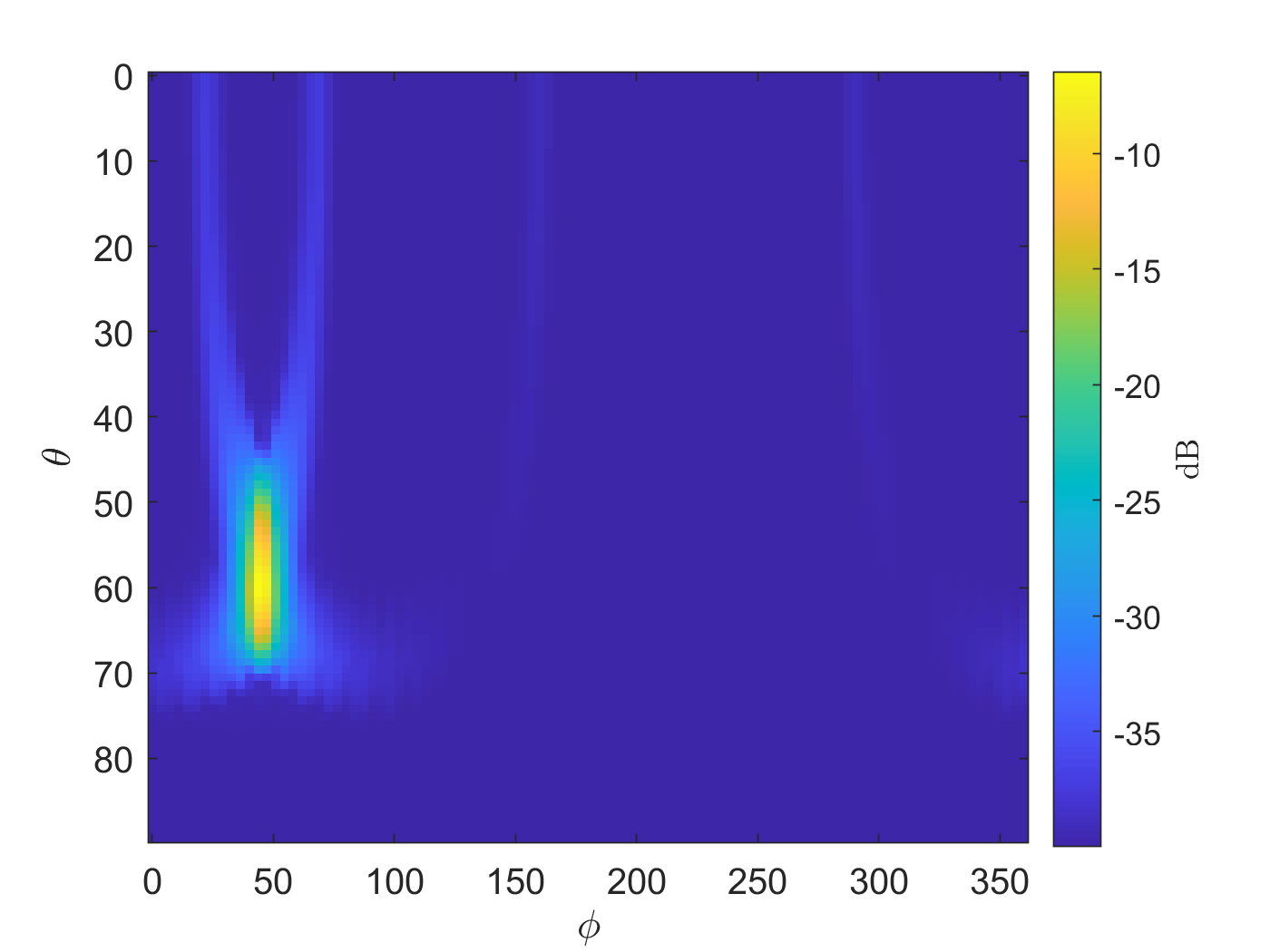}
            \caption%
            {{\small}}    
        \end{subfigure}
        
        % \begin{subfigure}[b]{0.45\textwidth}
        %     \centering
        %     \includegraphics[width=\textwidth]{figs/interferer_rejection_vs_K.png}
        %     \caption%
        %     {{\small}}    
        % \end{subfigure}
        % \begin{subfigure}[b]{0.45\textwidth}  
        %     \centering 
        %   \includegraphics[width=\textwidth]{figs/interferer_rejection_2D_1.png}
        %     \caption%
        %     {{\small}}    
        % \end{subfigure}
        
        \caption[Beam Pattern]
        {\small\sl MSE as a function of the interfering signals dynamic range for (a) a 256-element ULA steered to $45^o$, an interferer incident at $135^o$ and a variety of choices in $K$ and (b) a $32 \times 32$ element UPA steered to $\vtheta = [45^o,60^o]^T$, an interferer incident at $\vtheta = [225^o,60^o]^T$, and a variety of $K$. MSE given by \eqref{eq:interferer_MSE} as a function of the interfering signal DOA for the same (c) ULA and (d) UPA as previously described.} 
        \label{fig:interferer_rej}
\end{figure}

To illustrate this point we examine two array architectures: a 256-elements ULA and a $32\times32$ element UPA. The ULA is focused on a signal with $\gamma_0=1$ ,$f_c = 28$ GHz, $\Omega = 1.065$ GHz, and $\theta = 45^o$ such that the underlying subspace dimension is approximately 7. For the UPA we focus on a signal with $\gamma_0=1$, $f_c = 28$ GHz, $\Omega = 4.26$ GHz, and $\vtheta = [45^o,60^o]^T$ such that the underlying subspace dimension is 4. We assume a single interfering signal with the same respective center frequency and bandwidth is incident to the ULA and UPA at $135^o$ and $[225^o,60^o]^T$ respectively. We then vary the dynamic range $\gamma_1/\gamma_0$ and observe the MSE. Figure~\ref{fig:interferer_rej}(a) shows the result for a ULA while Figure~\ref{fig:interferer_rej}(b) displays the UPA result. In both cases an increase in $K$ decreases the MSE as expected. However, as the dynamic range is increased the gap between performance begins to close. Therefore though increasing $K$ will always reduce the MSE if the dynamic range between sources in substantial it may not be worth adding the additional measurements. 

When the signals are well separated in angular space the $K$ dominate eigenvectors of $\mR_\ell$ are approximately orthogonal to $\mV_K$ for $\ell>0$ \cite{Davenport:2011}. However, as the interferer's angle of arrival drifts closer to the steered direction this property no longer holds. Therefore the MSE has an angular dependence. To examine the affects of angle on the MSE we run a set of experiments similar to previous, but in this case we assume that $\gamma_1=\gamma_0$ and instead vary the DOA of the interfering signal. The results for the both the ULA and UPA are shown in Figure~\ref{fig:interferer_rej}(c-d) and as expected show the largest peak in the MSE coincides with when the interferer crosses over the signal of interest in angular space. Choosing $K$ to match the approximate subspace dimension makes the trailing eigenvalue sum (i.e. $\sum_{m=K+1}^{M}\gamma_0\lambda_m$) the dominate source of error outside the crossover region. Interestingly, when the sources are sufficiently spaced the interferer energy does not appear to appreciably affect the MSE until we choose $K$ to be significantly larger than $\ceil{2MW_{0}}$.

We note that the analysis and experiments in this section were performed with our now ``standard" choice in $\mPhi$. However, if the application demands a particular interferer (or more generally, a particular angle) to be nulled out then we can design a $\mPhi$ that is constrained to having this property. In particular, if we let $\mU_{K'}$ be the $K'$ dominant eigenvectors of $\mR_{I}$ then minimizing the MSE subject to the constraint $\norm{\mPhi\mU_{K'}}_F^2 \leq \delta$ for a tolerance parameter $\delta$ can produce the desired affect.

%--------------------------------------------------------------------------
% SECTION: Simplified measurement design

\section{Simplified measurement design}
\label{sec:Simplified_Measurements}

A key observation made in our reconstruction error estimates \eqref{eq:MSE_sol} in Section~\ref{sec:Measurement_and_recon} is that while the optimal choice for the measurement matrix $\mPhi$ is to match the eigenvectors of the covariance $\mR$, we can still have significant dimensionality reduction with other choices.  In this section, we describe a methodology for designing ``binary IQ'' $\mPhi$ that has entries 
%
%$\mPhi$ does not need to be comprised of transposed Slepian basis vectors. In fact we were able to broadband beamform signals using a drastically simplified architecture where 
$\Phi[m,n] \in \{1,-1,j,-j\}$.  This type of measurement system can be implemented with 2-bit resolution phase-shifters or purpose built hardware, making it feasible even at high frequencies.
%
%Though these binary IQ systems are extremely convenient for systems operating in restrictive environmental regimes the design of such a matrix is difficult. 
%
Our approach follows \cite{Duarte:2015,Duarte:2020}, which gives a principled design method for spectrally shaped binary sequences.

Given $\mR = \mV\mLambda\mV^\H$, we partition the eigenvectors as $\mV = \begin{bmatrix} \mV_K & \mV_\perp \end{bmatrix}$, where $\mV_K$ has the $K$ leading eigenvectors as columns and $\mV_\perp$ contains the eigenvectors corresponding to the $M-K$ smallest eigenvalues.
The problem amounts to designing the $K$ rows of $\mPhi$ such that its row space is as concentrated in the column space of $\mV_K$ as possible. 
%$\mathcal{R}(\mPhi^H)\approx \mathcal{R}(\mV_K)$. 
%
At the same time we would like to ensure that $\mPhi$ is reasonably well-conditioned, so that each of its rows measures something new. 
With $\{\vphi_k\}_{k=1}^K$ as the rows of $\mPhi$, one method to achieve this is to simultaneously minimize $\norm{\mV_{\perp}^H\vphi_k}_2^2$ and $|\inner{\mV_K^H\vphi_k,\mV_K^H\vphi_{k'}}|^2$ for $k'\neq k$.  We do this by designing the $\vphi_k$ iteratively, taking $\vphi_1$ as the solution to
\begin{align*}
    \minimize_{\vphi \in \mathbb{C}^M} \, \norm{\mV_{\perp}^H \vphi}_2^2 \quad \text{subject to} 
    &\quad \vphi[m] \in \{1,-1,j,-j\}~~\text{for}~~m = 1,\dots, M,
\end{align*}
and then for $k=2,\ldots,K$, taking $\vphi_k$ as the solution to
\begin{align*}
    \minimize_{\vphi \in \mathbb{C}^M} \, \norm{\mV_{\perp}^H \vphi}_2^2 \quad \text{subject to} 
    &\quad \vphi[m] \in \{1,-1,j,-j\}~~\text{for}~~m = 1,\dots, M, \\
    &\quad |\inner{ \mV_K^H \vphi,\mV_K^H \vphi_{k'}}|^2 \leq \alpha ~~\text{for}~~ k'=1,\ldots,k-1.
\end{align*} 
%where each iteration produces a new $\vphi_k$ that is subsequently incorporated into the first constraint for the next iteration.
This procedure finds the $\vphi_k$ that has most of its energy in the column space of $\mV_K$ while have bounded alignment (once projected into the column space) with the previous solutions. Of course an optimal choice would result in an orthogonal set of sequences, but this may be infeasible under the given constraints so we settle for a looser requirement with a threshold parameter $\alpha$. 

To cast this as an optimzation over the reals we let 
\begin{align*}
    \mU_\perp = 
    \begin{bmatrix}
     \Re{\mV_{\perp}^H}& -\Im{\mV_{\perp}^H}\\
     \Im{\mV_{\perp}^H} & \Re{\mV_{\perp}^H}
    \end{bmatrix}, ~ 
    \mU= 
    \begin{bmatrix}
     \Re{\mV_K^H}& -\Im{\mV_K^H}\\
     \Im{\mV_K^H} & \Re{\mV_K^H}
    \end{bmatrix}, ~
    \vh_k= 
    \begin{bmatrix}
     \Re{\vphi_k}\\
     \Im{\vphi_k} 
    \end{bmatrix}.
\end{align*}
Then the previous formulation can be expressed as an equivalent quadratically constrained quadratic program (QCQP)
\begin{align*}
    \minimize_{\vh_k \in \mathbb{R}^{2M}} \, \norm{\mU_\perp\vh_k}_2^2 \quad \text{subject to} 
    &\quad |\inner{ \mU\vh_k,\mU\vh_{k'}}|^2 \leq \alpha ~~ \text{for} ~~ k' < k\\
    &\quad \vh_k^2[m] = 1 ~~\text{for}~~ m = 1,\dots, 2M.
\end{align*}
We note that this will actually produce $\vphi_k[m] \in \{1+j,-1+j,1-j,-1-j\}$, but the objective and constraints are invariant under scalar multiplication. Hence we simply apply a constant phase shift to achieve the desired structure. 

This optimization problem is NP-Hard, and it is difficult to find good solutions in practice.  We can, however, use a standard convex relaxation to solve it approximately.  Using the facts that  $\norm{\mU_\perp\vh_k}_2^2 = \text{trace}(\mU_\perp^H \mU_\perp \vh_k \vh_k^H)$ and $|\inner{ \mU\vh_k,\mU\vh_{k'}}|^2 = \text{trace}(\mU^H \mU \vh_{k'} \vh_{k'}^H\mU^H \mU \vh_k \vh_k^H)$ we can then lift the vector variable $\vh_k$ to the matrix variable $\mT = \vh_k \vh_k^H $. We can then rewrite the QCQP as a conic optimization program over the semidefinite cone 
\begin{align*}
        \minimize_{\mT\in \mathbb{S}_{+}^{2M}} \, \text{trace}(\mU_\perp^H \mU_\perp \mT) \quad \text{subject to} &\quad \text{trace}(\mU^H \mU \vh_{k'} \vh_{k'}^H\mU^H \mU \mT) \leq \alpha \,\, \text{for} \, \, k' < k,\\
        &\quad \mT[m,m] = 1 \,\, \text{for}\, \, m = 1,\dots, 2M,\\
        &\quad \text{rank}(\mT) = 1,
\end{align*}
which identifies the rank constraint as the only non-convex portion of the program. The convex relaxation is to simply the rank constraint, thereby producing our final formulation 
\begin{align*}
        \minimize_{\mT\in \mathbb{S}_{+}^{2M}} \, \text{trace}(\mU_\perp^H \mU_\perp \mT) \quad \text{subject to} &\quad \text{trace}(\mU^H \mU \vh_{k'} \vh_{k'}^H\mU^H \mU \mT) \leq \alpha \,\, \text{for} \, \, k' < k,\\
        &\quad \mT[m,m] = 1 \,\, \text{for}\, \, m = 1,\dots, 2M.
\end{align*}
This is a standard semidefinite program that can be solved using well-established techniques. As there is no guarantee that the above produces a rank-1 solution or that the solution will factor into binary vectors, we need a method for projecting back onto the original constraint set.  This can be accomplished through a randomized search similar to that used in the MAXCUT algorithm \cite{Duarte:2015,Duarte:2020,Goemans:1995}: we draw $\vv \sim \mathcal{N}(\mtx{0},\mT)$ and generate a candidate solution vector $\hat{\vs}_k = \text{sign}(\vv)$. This is repeated for a fixed number of iterations and whichever candidate solution satisfies the constraints and has the smallest objective value is chosen as the (approximate) solution.

\begin{figure}[h]
        \centering
        \begin{subfigure}[b]{0.49\textwidth}
            \centering
            \includegraphics[width=\textwidth]{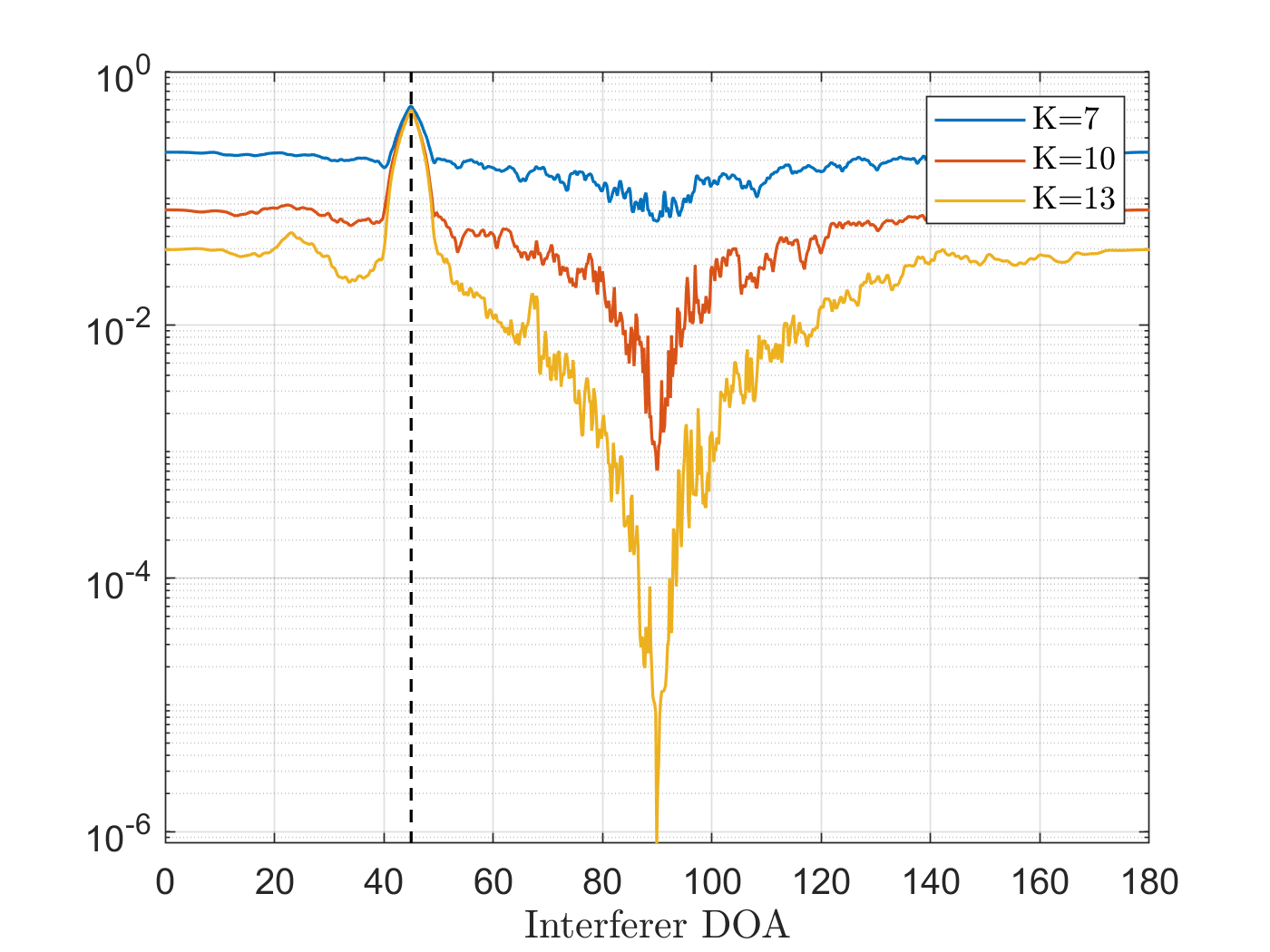}
            \caption%
            {{\small}}    
        \end{subfigure}
        \hfill
        \begin{subfigure}[b]{0.49\textwidth}  
            \centering 
            \includegraphics[width=\textwidth]{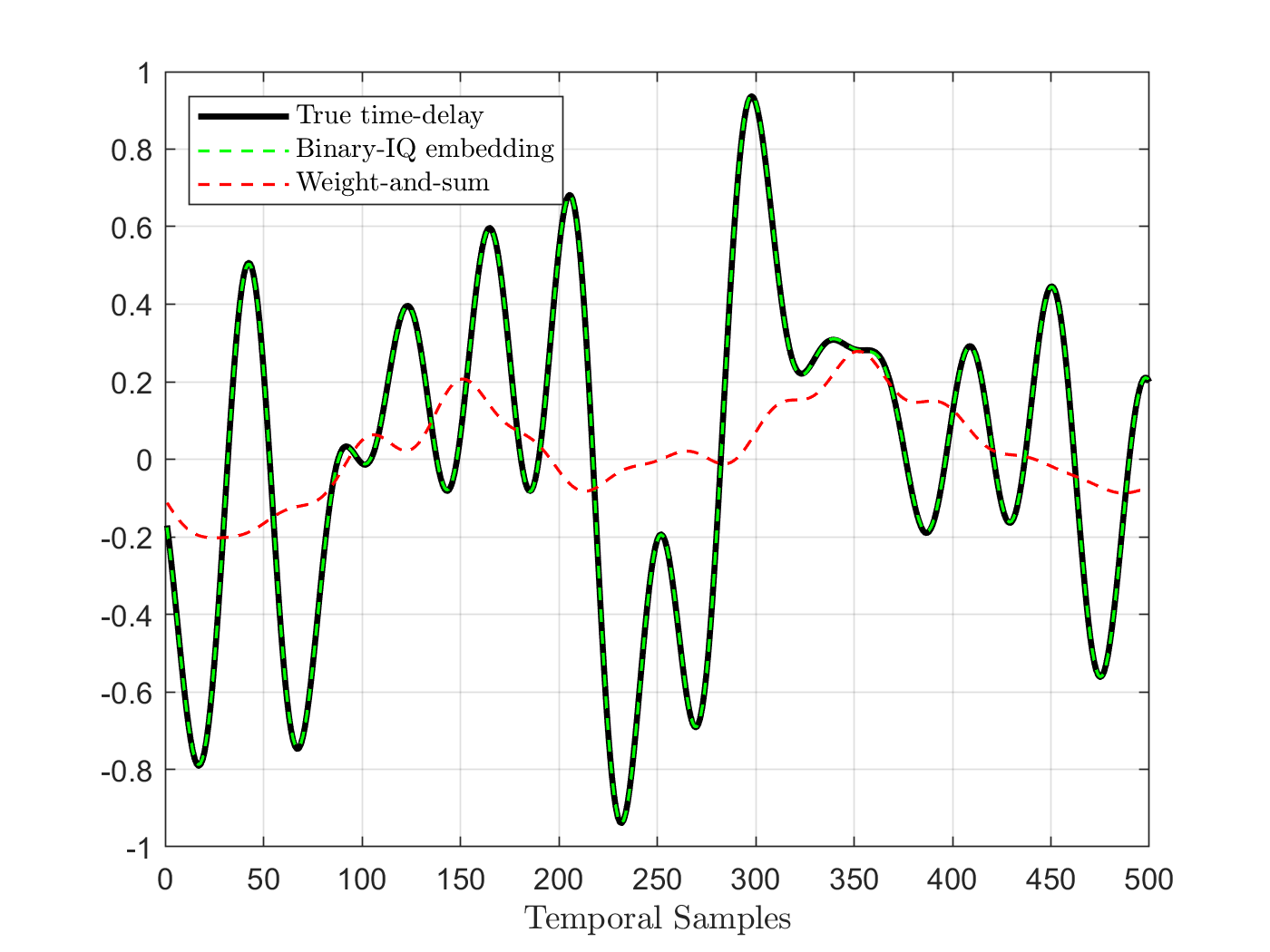}
            \caption%
            {{\small}}    
        \end{subfigure}
        \hfill
        \caption[Projection]
        {\small\sl (a) MSE as a function of interfering signal DOA for a 256-element ULA steered to $45^o$ and beamformed using varying number of $K$ binary IQ measurements. (b) Qualitative example of signal beamformed via $K=10$ binary IQ measurements compared to true-time delay and weight and sum.} 
        \label{fig:IQ_Beams}
\end{figure}

This algorithm was used to produce the binary IQ MSE plots in Figure~\ref{fig:Recovery}(c-d), which shows good recovery error despite the suboptimality of the measurements. To test how the binary IQ measurements fare in the presence of interfering signals we consider the same 256-element ULA scenario described in Section~\ref{sec:Signal_Isolation}. We generate a binary-IQ $\mPhi$ using the described algorithm and observe the MSE as the interfering signal sweeps across the array. The results shown in \ref{fig:IQ_Beams}(a) shows that MSE performance is generally worse than that achieved by the Slepian measurements. However, by oversampling (increasing $K$) we can achieve a reasonable level of performance. For instance setting $K=10$ essentially achieves the same level of performance as a set of 7 Slepian measurements. Of course the binary IQ measurements are far less spectrally concentrated than the Slepian basis vectors, and hence increasing $K$ does not improve performance at the same rate. As a more qualitative example we remove the interfering signal and estimate a series of $\vy_0$ given a set of binary IQ measurements. We then perform digital beamforming via multiple fractional-delay filters to produce a single output. Figure \ref{fig:IQ_Beams}(b) compares these results to a true-time delay and narrowband implementation, and as is apparent the transition to binary measurements results in no visually discernible loss in performance.
% To test this convex relaxation and randomized rounding scheme's ability to design the desired $\mPhi$ we test on an $M=128$ length sequence for a signal with a time-bandwidth product of 5. We set $\alpha = 10$ and let the randomized search produce $10^5$ candidate sequences. The resulting power spectrums of the $K=5$ measurement vectors are shown in Figure \ref{fig:IQ_Beams}a. Clearly the vectors show strong spectral concentration in the desired band, but there is a marked increase in sidelobe spurs compared to what would be seen with full resolution vectors. Figure \ref{fig:IQ_Beams}b shows the result of $\frac{1}{M} |\mPhi^H \mV_K\mV_k^H \mPhi|$ which demonstrates a strong diagonal behavior and subsequently offers good performance during the signal recovery stage of the beamformer. Overall this indicates the algorithm is able to produce suitable binary IQ beams for our proposed broadband beamforming scheme. As a reinforcement of this claim, this algorithm was used to produce the binary IQ MSE plots in Figure \ref{fig:Recovery}(c-d), where we saw good recovery despite the suboptimality of the measurements.

%--------------------------------------------------------------------------
\section{Conclusion}
\label{sec:Conclusion}

We have shown that by leveraging a robust Slepian subspace model it is possible to perform broadband beamforming in a manner distinct from traditional digital and true time delay beamforming. Furthermore, this method has the potential to significantly reduce the hardware complexity generally associated with broadband systems. Alongside formulating the method itself we have provided descriptions of fast Slepian techniques, temporal decimation schemes, signal isolation characteristics, and simplified measurement scheme designs. This supplementary material acts to bolster our argument that the system can be practically implemented at minimal hardware and computational cost. Ultimately we conclude that the proposed broadband beamforming technique may be a viable alternative to existing methods in a variety of applications.

%--------------------------------------------------------------------------

\bibliographystyle{IEEEtran}
\bibliography{refs}

\begin{thebibliography}{10}
\providecommand{\url}[1]{#1}
\csname url@samestyle\endcsname
\providecommand{\newblock}{\relax}
\providecommand{\bibinfo}[2]{#2}
\providecommand{\BIBentrySTDinterwordspacing}{\spaceskip=0pt\relax}
\providecommand{\BIBentryALTinterwordstretchfactor}{4}
\providecommand{\BIBentryALTinterwordspacing}{\spaceskip=\fontdimen2\font plus
\BIBentryALTinterwordstretchfactor\fontdimen3\font minus
  \fontdimen4\font\relax}
\providecommand{\BIBforeignlanguage}[2]{{%
\expandafter\ifx\csname l@#1\endcsname\relax
\typeout{** WARNING: IEEEtran.bst: No hyphenation pattern has been}%
\typeout{** loaded for the language `#1'. Using the pattern for}%
\typeout{** the default language instead.}%
\else
\language=\csname l@#1\endcsname
\fi
#2}}
\providecommand{\BIBdecl}{\relax}
\BIBdecl

\bibitem{SlepianI}
D.~Slepian and H.~O. Pollak, ``Prolate spheroidal wave functions, {F}ourier
  analysis and uncertainty — {I},'' \emph{Bell Systems Tech. J.}, vol.~40,
  no.~1, pp. 43--63, 1961.

\bibitem{SlepianII}
H.~J. Landau and H.~O. Pollak, ``Prolate spheroidal wave functions, {F}ourier
  analysis and uncertainty — {II},'' \emph{Bell Systems Tech. J.}, vol.~40,
  no.~1, pp. 65--84, 1961.

\bibitem{SlepianIII}
------, ``Prolate spheroidal wave functions, {F}ourier analysis and uncertainty
  - {III}: {T}he dimension of the space of essentially time- and band-limited
  signals,'' \emph{Bell Systems Tech. J.}, vol.~41, no.~4, pp. 1295--1336,
  1962.

\bibitem{SlepianVI}
D.~Slepian, ``Prolate spheroidal wave functions, {F}ourier analysis and
  uncertainty - {IV}: {E}xtensions to many dimensions; generalized prolate
  spheroidal functions,'' \emph{Bell Systems Tech. J.}, vol.~43, no.~6, pp.
  3009--3057, 1964.

\bibitem{SlepianV}
------, ``Prolate spheroidal wave functions, {F}ourier analysis, and
  uncertainty. {V} -- {T}he discrete case,'' \emph{Bell Systems Tech. J.},
  vol.~57, no.~5, pp. 1371--1430, 1978.

\bibitem{karnik2020improved}
S.~Karnik, J.~Romberg, and M.~A. Davenport, ``Improved bounds for the
  eigenvalues of prolate spheroidal wave functions and discrete prolate
  spheroidal sequences,'' \emph{Applied and Computational Harmonic Analysis},
  vol.~55, pp. 97--128, 2021.

\bibitem{karnik2017fast}
S.~Karnik, Z.~Zhu, M.~B. Wakin, J.~Romberg, and M.~A. Davenport, ``The fast
  slepian transform,'' \emph{Applied and Computational Harmonic Analysis},
  vol.~46, no.~3, pp. 624--652, 2019.

\bibitem{Karnik:2019:nonuniform}
S.~Karnik, J.~K. Romberg, and M.~A. Davenport, ``Bandlimited signal
  reconstruction from nonuniform samples,'' in \emph{Proc. Work. on Signal
  Processing with Adaptive Sparse Structured Representations (SPARS)}, 2019.

\bibitem{Schmidt:1986}
R.~Schmidt, ``Multiple emitter location and signal parameter estimation,''
  \emph{IEEE Trans. Antennas Propag.}, vol.~34, no.~3, pp. 276--280, 1986.

\bibitem{Spielman:1986}
D.~{Spielman}, A.~{Paulraj}, and T.~{Kailath}, ``Performance analysis of the
  music algorithm,'' in \emph{ICASSP '86. IEEE Int. Conf. on Acoust., Speech,
  and Signal Process}, vol.~11, 1986, pp. 1909--1912.

\bibitem{Buckley:1988:BASSALE}
K.~M. {Buckley} and L.~J. {Griffiths}, ``Broad-band signal-subspace
  spatial-spectrum ({BASS-ALE}) estimation,'' \emph{IEEE Trans. Acoust.,
  Speech, Signal Process}, vol.~36, no.~7, pp. 953--964, 1988.

\bibitem{Buckley:1986}
K.~{Buckley} and L.~{Griffiths}, ``Eigenstructure based broadband source
  location estimation,'' in \emph{ICASSP '86. IEEE Int. Conf. Acoust., Speech,
  Signal Process}, vol.~11, 1986, pp. 1869--1872.

\bibitem{Buckley:1987}
K.~{Buckley}, ``Spatial/spectral filtering with linearly constrained minimum
  variance beamformers,'' \emph{IEEE Trans. Acoust., Speech, Signal Process},
  vol.~35, no.~3, pp. 249--266, 1987.

\bibitem{Krim:1996}
H.~{Krim} and M.~{Viberg}, ``Two decades of array signal processing research:
  the parametric approach,'' \emph{IEEE Signal Processing Magazine}, vol.~13,
  no.~4, pp. 67--94, 1996.

\bibitem{anderson93op}
S.~Anderson, ``On optimal dimension reduction for sensor array signal
  processing,'' \emph{Signal Process.}, vol.~30, pp. 245--256, 1993.

\bibitem{Buckley:1987:RDBS}
X.~L. {Xu} and K.~M. {Buckley}, ``Reduced-dimension beam-space broad-band
  source localization: preprocessor design and evaluation,'' in \emph{Fourth
  Annual ASSP Workshop on Spectrum Estimation and Modeling}, 1988, pp. 22--27.

\bibitem{Ahmed:2017}
I.~{Ahmed}, H.~{Khammari}, A.~{Shahid}, A.~{Musa}, K.~S. {Kim}, E.~{De
  Poorter}, and I.~{Moerman}, ``A survey on hybrid beamforming techniques in
  {5G}: Architecture and system model perspectives,'' \emph{IEEE Commun.
  Surveys Tutorials}, vol.~20, no.~4, pp. 3060--3097, 2018.

\bibitem{Ali:2017}
E.~Ali, M.~Ismail, R.~Nordin, and N.~F. Abdullah, ``Beamforming techniques for
  massive {MIMO} systems in {5G}: overview, classification, and trends for
  future research,'' \emph{Frontiers of Information Technology \& Electronic
  Engineering}, vol.~18, pp. 753--772, 2017.

\bibitem{Bogale:2016}
T.~E. {Bogale}, L.~B. {Le}, A.~{Haghighat}, and L.~{Vandendorpe}, ``On the
  number of rf chains and phase shifters, and scheduling design with hybrid
  analog--digital beamforming,'' \emph{IEEE Trans. Wireless Commun.}, vol.~15,
  no.~5, pp. 3311--3326, 2016.

\bibitem{Sohrabi:2016}
F.~{Sohrabi} and W.~{Yu}, ``Hybrid digital and analog beamforming design for
  large-scale antenna arrays,'' \emph{IEEE J. Sel. Topics Signal Process.},
  vol.~10, no.~3, pp. 501--513, 2016.

\bibitem{Zhu_Li:2017}
D.~{Zhu}, B.~{Li}, and P.~{Liang}, ``A novel hybrid beamforming algorithm with
  unified analog beamforming by subspace construction based on partial csi for
  massive {MIMO-OFDM} systems,'' \emph{IEEE Trans. Commun}, vol.~65, no.~2, pp.
  594--607, 2017.

\bibitem{Jang:2019}
S.~{Jang}, R.~{Lu}, J.~{Jeong}, and M.~P. {Flynn}, ``A 1-{G}hz 16-element
  four-beam true-time-delay digital beamformer,'' \emph{IEEE J. Solid-State
  Circuits}, vol.~54, no.~5, pp. 1304--1314, 2019.

\bibitem{Chen:2018}
R.~{Chen}, H.~{Xu}, C.~{Li}, L.~{Zhu}, and J.~{Li}, ``Hybrid beamforming for
  broadband millimeter wave massive {MIMO} systems,'' in \emph{2018 IEEE 87th
  Vehicular Technology Conference (VTC Spring)}, 2018, pp. 1--5.

\bibitem{Peng:2018}
R.~{Peng} and Y.~{Tian}, ``Robust wide-beam analog beamforming with inaccurate
  channel angular information,'' \emph{IEEE Commun. Lett.}, vol.~22, no.~3, pp.
  638--641, 2018.

\bibitem{Morsali:2020}
A.~{Morsali} and B.~{Champagne}, ``Achieving fully-digital performance by
  hybrid analog/digital beamforming in wide-band massive-{MIMO} systems,'' in
  \emph{ICASSP 2020 - 2020 IEEE Int. Conf. on Acoust., Speech and Signal
  Processing (ICASSP)}, 2020, pp. 5125--5129.

\bibitem{Jung:2020}
M.~{Jung}, H.~J. {Yoon}, and B.~W. {Min}, ``A wideband true-time-delay phase
  shifter with 100\% fractional bandwidth using 28 nm {CMOS},'' in \emph{2020
  IEEE RFIC Symp.}, 2020, pp. 59--62.

\bibitem{Spoof:2020}
K.~{Spoof}, V.~{Unnikrishnan}, M.~{Zahra}, K.~{Stadius}, M.~{Kosunen}, and
  J.~{Ryyn{\"a}nen}, ``True-time-delay beamforming receiver with {RF}
  re-sampling,'' \emph{IEEE Trans. Circuits Syst. I, Reg. Papers}, pp. 1--13,
  2020.

\bibitem{Zatman:1998}
M.~{Zatman}, ``How narrow is narrowband?'' \emph{IEE Proceedings - Radar, Sonar
  and Navigation}, vol. 145, no.~2, pp. 85--91, 1998.

\bibitem{Davenport:2011}
M.~Davenport and M.~Wakin, ``Compressive sensing of analog signals using
  discrete prolate spheroidal sequences,'' \emph{Applied and Computational
  Harmonic Analysis}, vol.~33, 09 2011.

\bibitem{zhu18ro}
Z.~Zhu, S.~Karnik, M.~B. Wakin, M.~A. Davenport, and J.~Romberg, ``{ROAST}:
  {R}apid orthogonal approximate {S}lepian transform,'' \emph{IEEE Trans. Sig.
  Proc.}, vol.~66, no.~22, pp. 5887--5901, 2018.

\bibitem{Boulsane:2019}
M.~Boulsane, N.~Bourguiba, and A.~Karoui, ``Discrete prolate spheroidal wave
  functions: Further spectral analysis and some related applications,''
  \emph{Journal of Scientific Computing}, vol.~82, pp. 1--19, 2020.

\bibitem{Bonami:2021}
A.~Bonami, P.~Jaming, and A.~Karoui, ``Non-asymptotic behavior of the spectrum
  of the sinc-kernel operator and related applications,'' \emph{Journal of
  Mathematical Physics}, vol.~62, no.~3, p. 033511, Mar 2021.

\bibitem{Walden:1999}
R.~Walden, ``Analog-to-digital converter survey and analysis,'' \emph{IEEE J.
  Sel. Areas Commun.}, vol.~17, no.~4, pp. 539--550, 1999.

\bibitem{Hamam:2021}
T.~Hamam and J.~Romberg, ``Streaming solutions for time-varying optimization
  problems,'' \emph{arXiv}, 2021.

\bibitem{Keiner:2009}
J.~Keiner, S.~Kunis, and D.~Potts, ``Using {NFFT} 3---a software library for
  various nonequispaced fast fourier transforms,'' \emph{ACM Transactions on
  Mathematical Software}, vol.~36, no.~4, p.~19, 2009.

\bibitem{Duarte:2015}
D.~Mo and M.~F. Duarte, ``Design of spectrally shaped binary sequences via
  randomized convex relaxations,'' in \emph{2015 49th Asilomar Conf. on
  Signals, Systems and Computers}, 2015, pp. 164--168.

\bibitem{Duarte:2020}
------, ``Binary sequence set design for interferer rejection in multi-branch
  modulation,'' \emph{IEEE Trans. Signal Process.}, vol.~68, pp. 3769--3778,
  2020.

\bibitem{Goemans:1995}
M.~X. Goemans and D.~P. Williamson, ``Improved approximation algorithms for
  maximum cut and satisfiability problems using semidefinite programming,''
  \emph{Journal of the ACM}, vol.~42, no.~6, pp. 1115--1145, 1995.

\bibitem{mirsky75tr}
L.~Mirsky, ``A trace inequality of {John von Neumann},'' \emph{Monatshefte
  f\"ur Mathematik}, vol.~79, pp. 303--306, 1975.

\bibitem{horn12ma}
R.~A. Horn and C.~R. Johnson, \emph{Matrix Analysis}, 2nd~ed.\hskip 1em plus
  0.5em minus 0.4em\relax Cambridge University Press, 2012.

\end{thebibliography}

%--------------------------------------------------------------------------
\appendix

%--------------------------------------------------------------------------
\section{Optimal linear measurements}
\label{apx:optlinmeas}

Here we provide a quick argument that 
\begin{equation}
	\label{eq:maxtrace}
	\maximize_{\mPhi\in\mathbb{C}^{K\times M}} ~ \trace{\left(\mR\mPhi^\H(\mPhi\mR\mPhi^\H+\sigma^2\mId)^{-1}\mPhi\mR\right)} \quad\text{subject to}~~\|\mPhi\|\leq 1,
\end{equation}
is solved (and hence \eqref{eq:MSE} is minimized) when we take $\mPhi$ to be the $K$ leading eigenvectors of $\mR$.  This follows almost immediately from the von Neumann trace theorem \cite{mirsky75tr}, 
\cite[Chap.\ 7.4]{horn12ma} which states: if $\mA$ and $\mB$ are $M\times M$ conjugate symmetric positive semi-definite matrices\footnote{The theorem can be stated more generally in terms of the singular values of general matrices, but we will only need this special case below.} with eigenvalues $\lambda_1(\mA)\geq\lambda_2(\mA)\geq\cdots\geq\lambda_M(\mA)$ and $\lambda_1(\mB)\geq\lambda_2(\mB)\geq\cdots\geq\lambda_M(\mB)$, then
\[
	\trace(\mA\mB) \leq \sum_{m=1}^M\lambda_m(\mA)\lambda_m(\mB).
\]
If $\mA$ and $\mB$ have the same eigenvectors (when sorted based on the size of the corresponding eigenvalue), then equality above is achieved.

Let $\mR$ have eigenvalue decomposition $\mR = \mV\mLambda\mV^\H$ (where the elements along the diagonal are sorted from largest to smallest).  The search for $\mPhi$ can be recast as the search for its singular value decomposition components $\mPhi = \mU\mGamma\mW^\H$, where $\mU$ is $K\times K$ and orthonormal, $\mGamma$ is $K\times K$ with strictly positive entries down its diagonal, and $\mW$ is $M\times K$ with orthonormal columns.  Taking $\mZ = \mV^\H\mW$, we have 
\begin{align*}
	\trace(\mR\mPhi^\H(\mPhi\mR\mPhi^\H + \sigma^2\mId)^{-1}\mPhi\mR) 
	%&=\trace(\mLambda\mV^\T \mPhi^\T(\mPhi\mR\mPhi^\T + \sigma^2\mId)^{-1}\mPhi\mV\mLambda) \\
	&=\trace(\mV\mLambda\mZ\mGamma\mU^\H(\mU\mGamma\mZ^\H\mLambda\mZ\mGamma\mU^\H + \sigma^2\mId)^{-1}\mU\mGamma\mZ^\H\mLambda\mV) \\
	%&= \trace(\mLambda\mZ(\mZ^\T\mLambda\mZ + \sigma^2\mGamma^{-2})^{-1}\mZ^\T\mLambda) \\
	&=\trace(\mLambda^2\mZ(\mZ^\H\mLambda\mZ + \sigma^2\mGamma^{-2})^{-1}\mZ^\H).
\end{align*}
This is the trace of the product of two $M\times M$ symmetric semi-definite matrices $\mLambda^2$ and $\mZ(\mZ^\H\mLambda\mZ + \sigma^2\mGamma^{-2})^{-1}\mZ^\H$.  The latter has only $K$ non-zero eigenvalues, and we can induce their first $K$ eigenvectors to be the same by taking $\mZ =  \begin{bmatrix} \mId \\ \mzero \end{bmatrix}$, which happens when $\mW$ consists of the first $K$ columns of $\mV$, i.e.\  $\mW = \mV_K$.  With this choice of $\mW$, we have
\[
	\trace(\mR\mPhi^\H(\mPhi\mR\mPhi^\H + \sigma^2\mId)^{-1}\mPhi\mR) =
	\sum_{m=1}^K\lambda_m^2\cdot \frac{1}{\lambda_m + \sigma^2/\gamma_m^2},
\]
and we are left to choose the singular values $\{\gamma_m\}$ (note that there is no dependence on the left singular vectors $\mU$).  As $\lambda_m\geq 0 $ and $\sigma^2\geq 0$, the expression $(\lambda_m+\sigma^2/\gamma_m^2)^{-1}$ is monotonically increasing in $\gamma_m$, and so we want to choose each $\gamma_m$ as large as possible.  If we have the restriction $\|\mPhi\|\leq 1$, then it is optimal to take $\gamma_m=1$ for $m=1\ldots,K$.

Thus the solution to \eqref{eq:maxtrace} is $\mPhi = \mU\mV_K^\H$, where $\mU$ is any $M\times M$ orthonormal matrix.  In other words, $\mPhi^\H\mPhi$ is a projection onto the subspace spanned by the $K$ leading eigenvectors of $\mR$.

%--------------------------------------------------------------------------

%--------------------------------------------------------------------------

\section{MSE in the presence of interfering signals}
\label{apx:mse_large_K}

Here we show that term $\mPhi \mR_I \mPhi^H$ in \eqref{eq:interferer_MSE} behaves favorably when $\mPhi=\mV_K^H$ regardless of the choice in parameter $K$. To begin let $L=\sum_{\ell>0}\ceil{2MW_{\ell}}$ and $\mPhi = \mV_K^H$ for $K = \ceil{2MW_{0}}$. Given the eigen-decomposition $\mR_I = [\mU_L \ \mU_{\perp}]\mSigma[\mU_L \ \mU_{\perp}]^H$ we can write 
\begin{align*}
    \mPhi \mR_I \mPhi^H &= 
    \begin{bmatrix}
    \mV_K^H\mU_L & \mV_K^H\mU_{\perp}
    \end{bmatrix}
    \begin{bmatrix}
    \mSigma_L & \mtx{0}\\
    \mtx{0} & \mSigma_{\perp}
    \end{bmatrix}
    \begin{bmatrix}
    \mU_L^H\mV_K \\
    \mU_{\perp}^H\mV_{K} 
    \end{bmatrix} = \mV_K^H\mU_L\mSigma_L\mU_L^H\mV_K + \mV_K^H\mU_{\perp}\mSigma_{\perp}\mU_{\perp}^H\mV_{K}.
\end{align*}
The superposition of covariance matrices has a similar eigenvalue clustering behavior to the single interferer case such that $\norm{\mSigma_{\perp}}$ is very small regardless of the behavior of $\mV_K^H\mU_{\perp}$. Furthermore, assuming the signals are spectrally well separated then in general the columns of $\mV_K$ and $\mU_L$ are approximately orthogonal such that $\mV_K^H\mU_L\approx \mtx{0}$. Now lets begin adding an additional $K'$ set of measurements such that $\mPhi' = [\mV_K \ \mV_{K'}]^H$ so now 
\begin{align*}
    \mPhi' \mR_I \mPhi'^H &= 
    \begin{bmatrix}
    \mV_K^H\mU_L\mSigma_L\mU_L^H\mV_K & \mV_K^H\mU_L\mSigma_L\mU_L^H\mV_{K'}\\
    \mV_{K'}^H\mU_L\mSigma_L\mU_L^H\mV_{K} & \mV_{K'}^H\mU_L\mSigma_L\mU_L^H\mV_{K'}
    \end{bmatrix}
     + \mPhi' \mU_{\perp}\mSigma_{\perp}\mU_{\perp}^H \mPhi'^H,\\
     &\preceq
     \begin{bmatrix}
    \mV_K^H\mU_L\mSigma_L\mU_L^H\mV_K & \mV_K^H\mU_L\mSigma_L\mU_L^H\mV_{K'}\\
    \mV_{K'}^H\mU_L\mSigma_L\mU_L^H\mV_{K} & \mV_{K'}^H\mU_L\mSigma_L\mU_L^H\mV_{K'}
    \end{bmatrix}
     + \tilde{\epsilon}_1\mtx{I},\\
     &\approx
     \begin{bmatrix}
    \mtx{0} & \mtx{0}\\
    \mtx{0} & \mV_{K'}^H\mU_L\mSigma_L\mU_L^H\mV_{K'}
    \end{bmatrix}
     + \tilde{\epsilon}_1\mtx{I}.\\
    %  \begin{bmatrix}
    % \mV_K^H\tilde{\mV}_{\perp}\tilde{\mLambda}_{\perp}\tilde{\mV}_{\perp}^H\mV_K & \mV_K^H\tilde{\mV}_{\perp}\tilde{\mLambda}_{\perp}\tilde{\mV}_{\perp}^H\mV_{K'}\\
    % \mV_{K'}^H\tilde{\mV}_{\perp}\tilde{\mLambda}_{\perp}\tilde{\mV}_{\perp}^H\mV_{K} & \mV_{K'}^H\tilde{\mV}_{\perp}\tilde{\mLambda}_{\perp}\tilde{\mV}_{\perp}^H\mV_{K'}
    % \end{bmatrix}.\\
\end{align*}
So we have a matrix that is generally small regardless of our choice in $K$ or $K'$ and a matrix that is approximately zero everywhere except its lower diagonal. Letting $\mZ = \mLambda_{K'}+(\tilde\epsilon_1 + \sigma^2)\mtx{I} + \mV_{K'}^H\mU_L\mSigma_L\mU_L^H\mV_{K'}$ then
\begin{align*}
    \mathbb{E}_{\vy_0,\dots,\vy_{L-1},\veta}\left\{\norm{\hat\vy-\vy_0}_2^2 \big | \vw \right\} & \lessapprox
    \sum_{m=1}^K \gamma_0 \lambda_m \left ( 1 - \frac{\gamma_0\lambda_m}{\gamma_0\lambda_m +\tilde{\epsilon}_1+ \sigma^2} \right) + \trace{ \left ( \mLambda_{K'}-\mLambda_{K'}\mZ^{-1}\mLambda_{K'}\right )}\\ & \quad + \sum_{m=K+K'+1}^M \gamma_0 \lambda_m,
\end{align*}
so even when $\norm{\mZ}$ is large it does not affect the MSE to a great extant since the trace term is generally small due to the eigenvalue clustering\footnote{This follows from the fact $\trace{(\mLambda_{K'}-\mLambda_{K'}\mZ^{-1}\mLambda_{K'})} \leq \trace{(\mLambda_{K'})}$ and $\trace{(\mLambda_{K'})}$ is generally small.}. Of course $\mLambda_{K'} \succeq \mLambda_{K'}-\mLambda_{K'}\mZ^{-1}\mLambda_{K'}$ so the MSE will never increase with an increase in the number of measurements. However, the degree to which this improves the MSE depends largely on $\mZ$. In summary, we expect that the addition of an interferer will most significantly affect the reconstruction along the lower variance principle axis of $\mR_0$ where a smaller portion of the energy in $\vy_0$ is contained.

The above discussion is meant to establish a very general framework for why the term $\mPhi\mR_{I}\mPhi^H$ does not affect the dominant terms in the MSE expression to a great extent. If we make further assumptions on the array this argument can be made far more precise. Consider a ULA and a ``worst case" interferer that occupies the entire spectrum outside the signal of interest's band such that $\mR_I \preceq \gamma_{max}(\mtx{I}-\gamma_0^{-1}\mR_0)$ where $\gamma_{max} = \max_{\ell>0}\gamma_\ell$ and $\mR_0 = \gamma_0\mV\mLambda\mV^H$. In this case $\mV_K = \mU_{\perp}$ such that 
\begin{align*}
    \mPhi' \mR_I \mPhi^{'H} &= 
    \begin{bmatrix}
    \mtx{0} & \mtx{0}\\
    \mtx{0} & \mV_{K'}^H\mU_L\mSigma_L\mU_L^H\mV_{K'}
    \end{bmatrix} + 
     \begin{bmatrix}
    \mSigma_{\perp} & \mtx{0}\\
    \mtx{0} & \mtx{0}
    \end{bmatrix}.\\
\end{align*}
From here we have $\mSigma_{\perp} = \gamma_{max}(\mtx{I}_K-\mLambda_K)$ and $\mV_{K'}^H\mU_L\mSigma_L\mU_L^H\mV_{K'}= \gamma_{max}(\mtx{I}_{K'}-\mLambda_{K'})$ and can subsequently bound the MSE by
\begin{align*}
    \mathbb{E}_{\vy_0,\dots,\vy_{L-1},\veta}\left\{\norm{\hat\vy-\vy_0}_2^2 \big | \vw \right\} & \leq 
    \sum_{m=1}^{K+K'} \gamma_0 \lambda_m \left ( 1 - \frac{\gamma_0\lambda_m}{\gamma_0\lambda_m +\gamma_{max}(1-\lambda_m)+ \sigma^2} \right) + \sum_{m=K+K'+1}^M \gamma_0 \lambda_m.
\end{align*}
We note that for $L'=\ceil{2MW_0}$ the $k$th largest eigenvalues satisfy 
\begin{align}
    \label{eq:lambdak_plateau1}
    \lambda_k \geq 1 - c_5 \exp\left( \frac{k-L'}{c_6} \right), \quad k\leq L',
\end{align}
where $c_5$ and $c_6$ are known and reasonably small constants \footnote{See \cite[Cor.\ 1]{karnik2020improved} for a precise statement of \eqref{eq:lambdak_plateau1}.}. So the eigenvalues rapidly approach 1 as we recede from the edge of the plateau at $k=L'$. Therefore the term $1-\lambda_m$ will be very close to zero for $m = 1,\dots,K$ but may be substantially larger than 0 for $m = K+1,\dots,K+K'$. If we consider the case of a large dynamic range disparity where $\gamma_0 \ll \gamma_1$ then the $m = K+1,\dots,K+K'$ terms will become highly damped such that increasing $K'$ past a certain point has a diminished effect on lowering the MSE.

%--------------------------------------------------------------------------
\end{document}